\documentclass[aps,prd,reprint,superscriptaddress]{revtex4-1}
\pdfoutput=1
\usepackage[T1]{fontenc}
\usepackage{setspace}
\usepackage{amsmath}
\usepackage[svgnames]{xcolor}
\usepackage{graphicx}
\usepackage{placeins}
\usepackage{units}
\usepackage{xspace}
\newcommand{\mb}{MiniBooNE\xspace}
\newcommand{\minerva}{MINER\ensuremath{\nu}A\xspace}

\newcommand{\ccqe}{CCQE\xspace}
\newcommand{\qq}{\ensuremath{Q^{2}}\xspace}
\newcommand{\qqqe}{\ensuremath{Q^{2}_{\textrm{QE}}}\xspace}
\newcommand{\pf}{\ensuremath{p_{\mathrm{F}}}\xspace}
\newcommand{\eb}{\ensuremath{E_{\mathrm{b}}}\xspace}
\newcommand{\ma}{\ensuremath{M_{\textrm{A}}}\xspace}
\newcommand{\Enu}{\ensuremath{E_{\nu}}\xspace}
\newcommand{\enuqe}{\ensuremath{E_{\nu}^{\mbox{\scriptsize{QE,\;FG}}}}\xspace}

\newcommand{\neut}{\textsc{NEUT}\xspace}
\usepackage{tabulary}
\usepackage{caption}
\usepackage{subcaption}
\usepackage{multirow}
\begin{document}

\title{Direct extraction of nuclear effects in quasielastic scattering on carbon}
\date{\today}
\author{Callum Wilkinson}
\affiliation{University of Bern, Albert Einstein Center for Fundamental Physics, Laboratory for High Energy Physics (LHEP), Bern, Switzerland}
\author{Kevin S.\ McFarland}
\affiliation{University of Rochester, Department of Physics and Astronomy, Rochester, New York 14627 USA}
\begin{abstract}

Nuclear effects on neutrino reactions are expected to be a significant complication in current and future neutrino oscillation experiments seeking precision measurements of neutrino flavor transitions.  Calculations of these nuclear effects are hampered by a lack of experimental data comparing neutrino reactions on free nucleons to neutrino reactions on nuclei.  We present results from a novel technique that compares neutrino and antineutrino charged current quasielastic scattering on hydrocarbons to extract a cross section ratio of antineutrino charged current elastic reactions on free protons to charged current quasielastic reactions on the protons bound in a carbon nucleus.  This measurement of nuclear effects is compared to models. 

\end{abstract}
\maketitle

The cross sections for neutrino and antineutrino charged current
quasielastic (CCQE) reactions on free nucleons, $\nu_\ell n\to\ell^-
p$ and $\bar{\nu}_\ell p\to\ell^+ n$, can be expressed in terms 
of nucleon form factors~\cite{adler63,marshak69,pais71,smith72}.
This prescription, with form factors constrained by electron nucleon
elastic scattering and pion electroproduction data, accurately
describes available neutrino interaction data on hydrogen and loosely
bound deuterium
targets~\cite{bba03,kuzmin_2006,bbba07,kuzmin_2008,Bodek:2007vi}.  On
heavier, more tightly bound nuclei, the relativistic Fermi Gas (FG)
model~\cite{smith-moniz72} modifies this formalism within the context
of the impulse approximation to include a simple description of the
initial state of bound nucleons within the nucleus, and has been
extensively used in neutrino interaction generators.  However, 
experiments with carbon, oxygen and iron
targets~\cite{mbCCQE, k2kCCQE, NOMAD_CCQE, minosCCQE,
  minerva-nu-ccqe, minerva-antinu-ccqe, mbAntiCCQE, ingridCCQE,
  t2kCCQE}
 with neutrino energies of a few GeV have measured a significantly different, typically higher, quasielastic cross section than predicted by the FG model. Additionally, recent measurements of the CC-inclusive cross section have shown that nuclear effects are not well understood~\cite{Rodrigues:2015hik}, and that the ratio of CC-inclusive cross section measurements on different nuclear targets cannot be described by the models available in generators~\cite{minerva_ccinc_2014}, particularly in the elastic region.

Theoretical work to understand these differences have been focused on
three broad areas: a more sophisticated description of the initial
state of nucleons within the nucleus~\cite{sf, ankowski_SF,
  butkevich_2009, eff-sf, leitner_2009, madrid_2003, meucci_2003,
  pandey_2014}; contributions to the cross section beyond the impulse
approximation which involve multiple initial state nucleons (hereafter
referred to as multinucleon processes or MNP)~\cite{nieves_2011,
  martini_2009}; and collective effects which modify the cross
section, which are generally referred to by the name of the
calculation, the Random Phase Approximation (RPA)~\cite{nieves_2011,
  martini_2009}. Despite the flurry of theoretical activity in recent
years, a consistent picture has yet to emerge, in part because of
significant differences in the predictions of theoretical
calculations~\cite{zeller12, hayato_review_2014, garvey_review_2014}.

Quasielastic interactions are especially important for accelerator
neutrino oscillation experiments at GeV
energies~\cite{mb_nue_app_2009,t2k_nue_2011,nova, hyperk, dune}.
In the impulse approximation, the initial state nucleons are
independent in the mean field of the nucleus, and therefore the
neutrino energy and momentum transfer $Q^2$ can be estimated from the
polar angle $\theta_\ell$ and momentum $p_\ell$ of the final state
lepton.  However, the initial state prescription and multinucleon
processes both disrupt this relationship in different
ways~\cite{Martini:2012uc,Lalakulich:2012hs,Nieves:2012yz}. MNP
and collective RPA processes both alter the distribution of $Q^2$
which can in turn alter the relative acceptance of near and far
detectors.  Therefore understanding nuclear modifications is essential
for the current and future generations of neutrino oscillation
experiments.

Although neutrino--nucleon scattering data would be invaluable for untangling nuclear effects, no new data are expected from any current or planned experiments in the few-GeV energy region. In this analysis, we present a method for extracting a measurement of
the suppression and enhancement to the CCQE cross section due to nuclear effects in
carbon from neutrino and antineutrino measurements on hydrocarbon
targets, which is relatively free of axial form factor and other uncertainties, particularly at low \qq. This method is largely model independent when applied to
high energy CCQE data, such as that from
\minerva~\cite{minerva-nu-ccqe, minerva-antinu-ccqe}, but less so at
the lower energies of the \mb experiment~\cite{mbCCQE, mbAntiCCQE}.

The \ccqe neutrino--nucleon differential cross section for free
nucleons as a function of the negative of the four-momentum transfer
squared, $\qq$, can be expressed using the Llewellyn-Smith
formula~\cite{smith72}:
\begin{align}
\frac{d\sigma}{d\qq} &{\nu_{l}n \rightarrow l^{-}p \choose \bar{\nu}_{l}p \rightarrow l^{+}n} = \frac{M^2G_{\mathrm{F}}^2\cos^2\vartheta_C}{8\pi \Enu^2}\notag\\\vspace{6pt}
&\times\left[ A(\qq)\pm B^\prime(\qq)\frac{(s-u)}{M^4} + C(\qq)\frac{(s-u)^2}{M^4} \right],\label{eq:smith_xsec}
\end{align}
\noindent where $M$ is the mass of the struck nucleon,
$G_{\mathrm{F}}$ is Fermi's constant, $\vartheta_C$ is the Cabibbo
angle, $E_\nu$ is the incoming neutrino energy and $s$ and $u$ are the Mandelstam variables. $A(\qq)$, $B^\prime(\qq)$ and $C(\qq)$ are functions of the vector form factors: $F_{\mathrm{V}}^{1,\; 2}$, constrained by electron nucleon elastic scattering experiments~\cite{bba03,bbba07}; the axial form factor, $F_{\mathrm{A}}$, constrained by neutrino scattering experiments on hydrogen and deuterium and from pion electroproduction~\cite{kuzmin_2006,bbba07,kuzmin_2008,Bodek:2007vi}; and the pseudoscalar form factor, $F_{\mathrm{P}}$, which is derived from $F_{\mathrm{A}}$~\cite{smith72}. Uncertainties from $F_{\mathrm{P}}$ and the assumption that second class currents can be neglected are discussed in Reference~\cite{day_nue_numu_2012}.
The term with $B^\prime(\qq)$ contains the interference between the axial and vector
currents, and it is this term which is responsible for the $\qq$
dependent difference between the $\nu_\ell + n \rightarrow \ell^{-} + p$
and $\bar{\nu}_\ell + p \rightarrow \ell^{+} + n$ cross sections.  At $\qq
= 0$, there is no difference between the \ccqe cross sections for
neutrinos and antineutrinos.  Note that $s-u=4ME_\nu-Q^2-m_\ell^2$, where
$m_\ell$ is the mass of the final state lepton; therefore, the effect of
the interference term is largest at small neutrino energies and high $Q^2$.

Nuclear models available in the NEUT~\cite{neut, niwg_paper} event generator will be compared to the data.  NEUT's default model is the
Smith-Moniz~\cite{smith-moniz72} implementation of an FG model with
Fermi momentum (\pf) and binding energy (\eb) on carbon set to 
 \pf = 217 MeV and \eb = 25 MeV based on electron scattering
data~\cite{moniz_eA_RFG_1971}.  NEUT has implemented the Spectral
Function (SF) model of Benhar~\cite{sf,andy_thesis} which describes the initial
nucleon's correlated momentum and removal energy and includes short
range nuclear correlations which affect $\sim$20\% of the CCQE rate.
Nuclear screening due to long-range nucleon correlations is implemented in 
RPA calculations~\cite{nieves_2011}.  Calculations of
MNP use the model of Nieves~{\it et al.}~\cite{nieves_2011, nievesExtension}.
NEUT also has implementations of two effective models constructed to
ensure agreement with electron data, an Effective Spectral Function
(ESF)~\cite{eff-sf,eff-sf_proc,my_thesis} and the Transverse
Enhancement Model (TEM)~\cite{tem,my_thesis}.   
For all models, we use the BBBA05 vector nucleon form
factors~\cite{bbba05} and a dipole axial form factor with \ma = 1.00
GeV, based on fits to bubble chamber data~\cite{kuzmin_2006,bbba07,kuzmin_2008,Bodek:2007vi}. 

In this analysis we use the published flux-averaged neutrino and
antineutrino CCQE cross section results on hydrocarbon targets from
the \minerva~\cite{minerva-nu-ccqe, minerva-antinu-ccqe} and
\mb~\cite{mbCCQE, mbAntiCCQE} experiments. The results used are differential in terms of \qqqe, 
derived from lepton kinematics under the
quasielastic hypothesis,
\begin{gather}
  \qqqe = -m_{\mu}^{2}+2\enuqe(E_{\mu}-\sqrt{E_{\mu}^{2}-m_{\mu}^{2}}\cos{\theta_{\mu}}),\notag\\
  \enuqe = \frac{2M'_{i}E_{\mu}-(M'^{2}_{i}+m_{\mu}^{2}-M^{2}_{f})}{2(M'_{i}-E_{\mu}+\sqrt{E_{\mu}^{2}-m_{\mu}^{2}}\cos{\theta_{\mu}})}, \label{eq:ccqe-q2qe}
\end{gather}
where $E_\mu$ is the muon energy, $m_{\mu}$ is the muon mass, $M_i$ ($M_f$) is the initial (final) nucleon mass, and
$M'_{i} = M_{i} - V$ where $V$ is the effective binding energy. For both \mb datasets and for the \minerva neutrino dataset, $V = 34$
MeV; for the \minerva antineutrino dataset, $V = 30$ MeV.

\begin{figure}[htbp]
  \centering
  \includegraphics[width=0.45\textwidth]{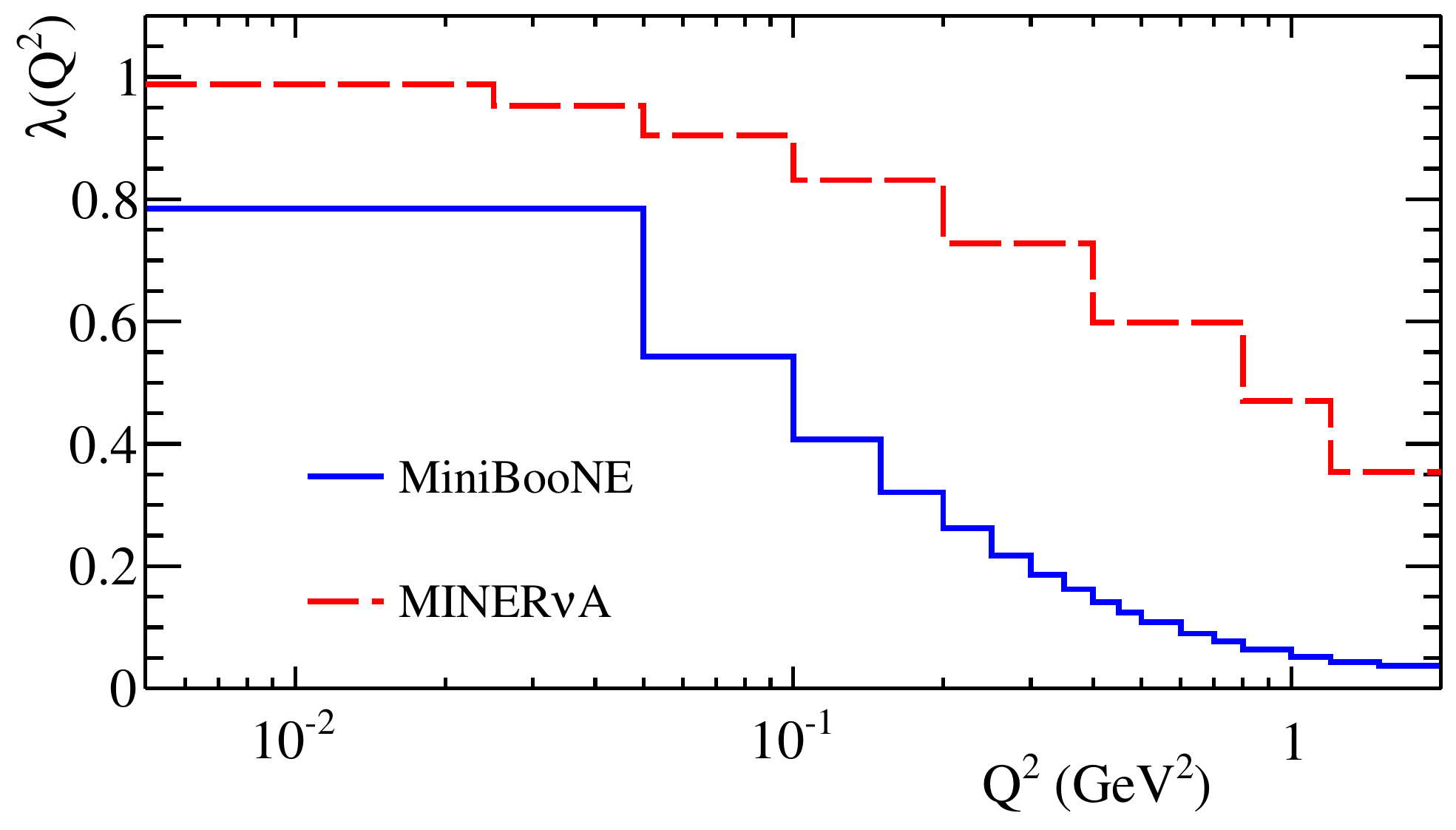}
  \caption{$\lambda(\qq) = \sigma^{\bar{\nu}}_{p}(\qq)/\sigma^{\nu}_{n}(\qq)$ calculated using the free nucleon cross-sections implemented in the GENIE neutrino interaction generator~\cite{genieMC}, averaged over the relevant flux and binned into the $\qq$ binning used by the relevant experiment. The values are given in Appendix~\ref{appendix:applying}.}\label{fig:lambda}
\end{figure}

There are three differences in the neutrino and antineutrino cross
section measurements for CCQE-like processes on hydrocarbon, $CH_{\mathrm{N}}$
targets. Firstly, the
neutrino and antineutrino cross sections are fundamentally different for free
nucleons (see Equation~\ref{eq:smith_xsec}). Secondly, the neutrino and antineutrino fluxes produced in
the same beamline may be different~\cite{numi-beam, mb-flux}. Finally,
antineutrinos can interact with the free proton from the hydrogen as
well as bound protons within the carbon nucleus, whereas neutrinos can
only interact with bound neutrons. The central thesis of this work is that a direct measurement of nuclear
effects in carbon can be made by
\begin{equation}
  \label{eq:ts}
  \frac{6\sigma^{\bar{\nu}}_{\mathrm{H}}}{\sigma_{\mathrm{C}}^{\bar{\nu}}} = \frac{\left[(6+N)\widetilde{\sigma}^{\bar{\nu}}_{\mathrm{CH_N}} - 6 \lambda(Q^2)\widetilde{\sigma}^{\nu}_{\mathrm{CH_N}}\right]}{N\lambda(Q^2) \widetilde{\sigma}^{\nu}_{\mathrm{CH_N}}},
\end{equation}
where $\sigma$ denotes the flux-averaged cross section for
interactions between the neutrino species in the superscript and the
the target in the subscript; $\widetilde{\sigma}$ denotes a cross
section per nucleon; the correction factor $\lambda(Q^2) =
(d\sigma^{\bar{\nu}}_{p}/dQ^2)/(d\sigma^{\nu}_{n}/dQ^2)$ corrects for
the difference between the neutrino and antineutrino nucleon cross
sections and fluxes and is shown in Figure~\ref{fig:lambda}.

\begin{figure}[h!]
  \centering
  \begin{subfigure}{0.45\textwidth}
    \includegraphics[width=\textwidth]{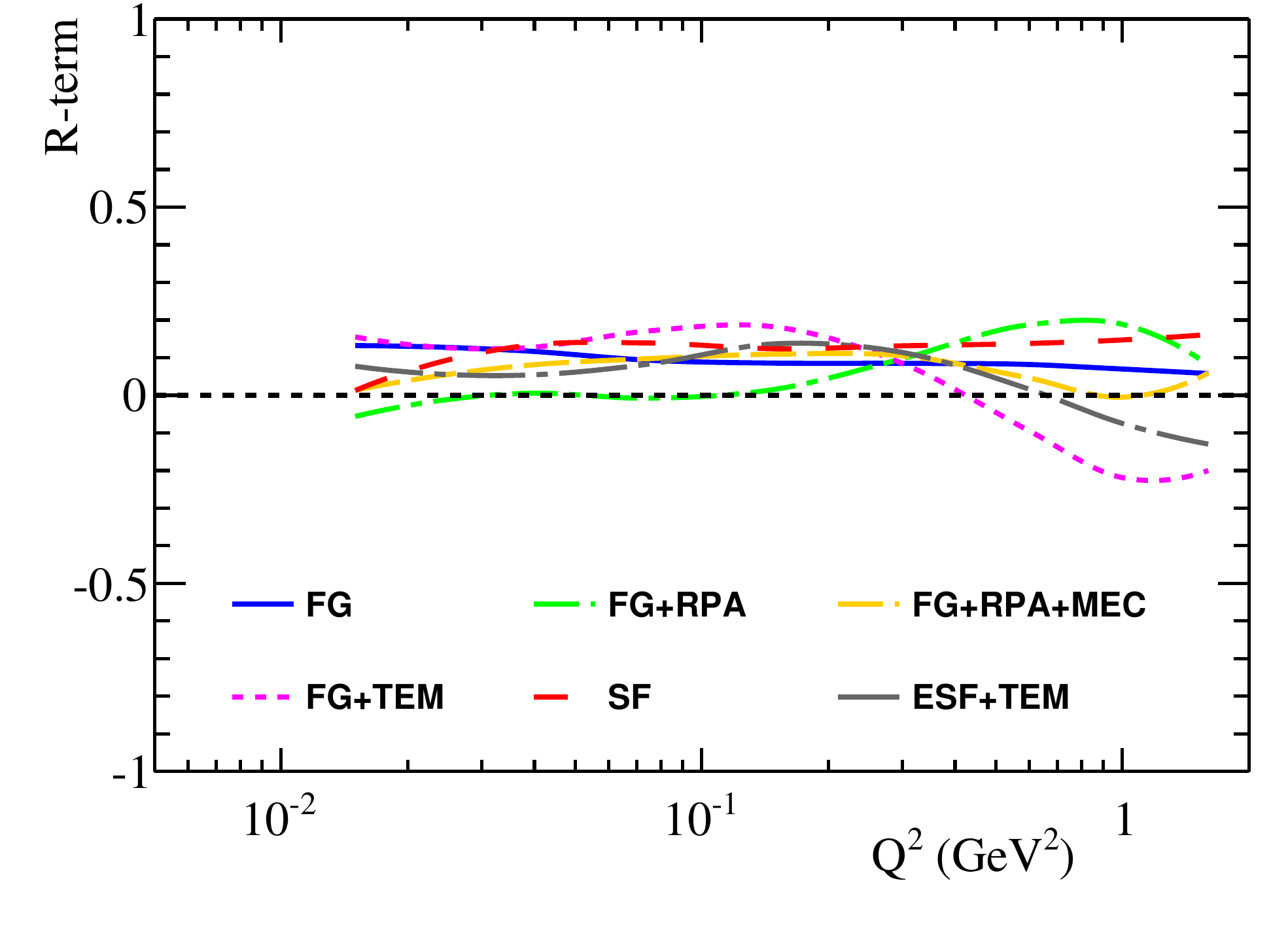}
    \caption{\minerva}
    \label{fig:r_minerva}
  \end{subfigure}
  \begin{subfigure}{0.45\textwidth}
    \includegraphics[width=\textwidth]{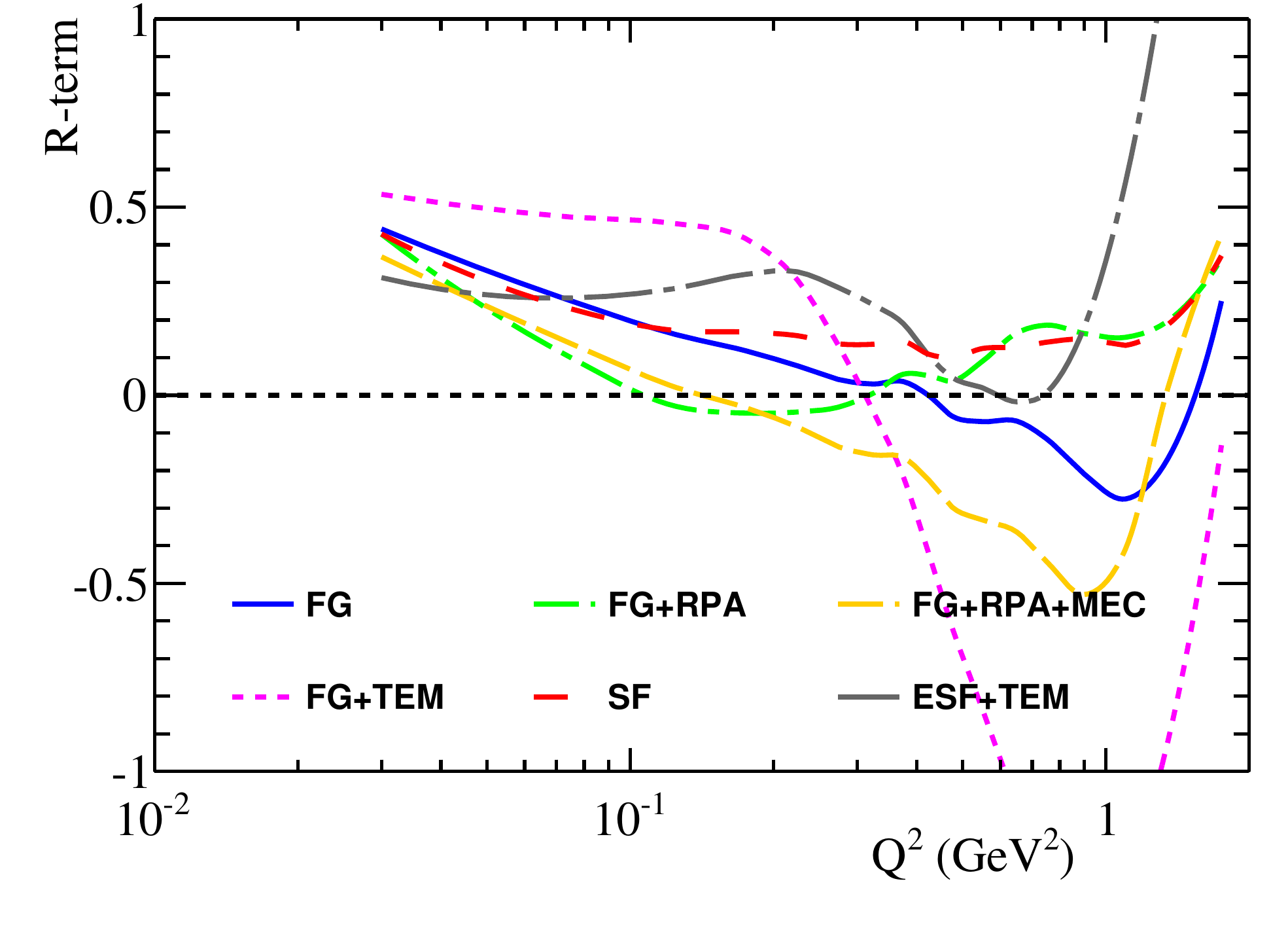}
    \caption{\mb}
    \label{fig:r_mb}
    \end{subfigure}
  \caption{The $R$-term, defined in Equation~\ref{eq:ts-r}, is shown for both \minerva and \mb, for a variety of models. It shows the size of the bias on the value $\frac{6\sigma^{\bar{\nu}}_{\mathrm{H}}}{\sigma_{\mathrm{C}}^{\bar{\nu}}}$, extracted using Equation~\ref{eq:ts}, which is due to our assumption that the neutrino and antineutrino cross section ratio is the same for free nucleons and bound nucleons. A value of 0 indicates no bias. The statistical error from the MC is $\sim$0.05 for all bins and is uncorrelated between all bins and models.}\label{fig:r_ratios}
\end{figure}

The validity of Equation~\ref{eq:ts} rests on the
assumption that the ratio of bound to free cross sections, as a function of \qq, is the
same for neutrino and antineutrino scattering. The quality of this assumption can be tested directly for a variety of models by looking at the double ratio $R(\qq)$,
\begin{equation}
  R(\qq) = \left.\left(\frac{6\sigma^{\bar{\nu}}_{p}(\qq)}{\sigma^{\bar{\nu}}_{C}(\qq)}\right) \middle/ \left(\frac{6\sigma^{\nu}_{n}(\qq)}{\sigma^{\nu}_{C}(\qq)}\right)\right.,
  \label{eq:r}
\end{equation}
where the bound CCQE cross section for neutrino and antineutrino ($\sigma^{\bar{\nu}}_{C}(\qq)$ and $\sigma^{\nu}_{C}(\qq)$) is calculated for any given nuclear model. 
Deviations of $R$ from 1 indicate that this assumption is inadequate and will lead to biases in results extracted with Equation~\ref{eq:ts}. Within an FG model, the assumption that $R=1$ is imperfect due to the effects of binding energy and kinematic boundaries, and this point is discussed further in Appendix~\ref{appendix:nu-nubar-equality}.
The bias to our extracted results can be seen in the generalization of Equation~\ref{eq:ts} for the case where $R\ne 1$:
\begin{eqnarray}
  \label{eq:ts-r}
  \frac{6\sigma^{\bar{\nu}}_{\mathrm{H}}}{\sigma_{\mathrm{C}}^{\bar{\nu}}} &=& \frac{\left[(6+N)\widetilde{\sigma}^{\bar{\nu}}_{\mathrm{CH_N}} - 6 \lambda(Q^2)\widetilde{\sigma}^{\nu}_{\mathrm{CH_N}}\right]}{N\lambda(Q^2) \widetilde{\sigma}^{\nu}_{\mathrm{CH_N}}}, \nonumber\\
&&+\left[\frac{6+N}{N\lambda(Q^2)}\frac{\widetilde{\sigma}^{\bar{\nu}}_{\mathrm{CH_N}}}{\widetilde{\sigma}^{\nu}_{\mathrm{CH_N}}}(R-1)\right] \rightarrow \mathrm{{\it R}-term}.
\end{eqnarray}
We determine the size of the $R$-term \minerva and \mb fluxes for the nuclear models discussed above in Figure~\ref{fig:r_ratios}. The $R$-term is relatively flat across the entire \qq range for \minerva, with no indication of strong biases, which suggests that our assumption holds well in this case and our results will be unbiased and do not depend strongly on the choice of nuclear model. For \mb, the assumption does not hold up as well, so we expect biases in results extracted using Equation~\ref{eq:ts}.

Another complication of this analysis is that experiments measure
differential cross-sections in \qqqe, as defined in
Equation~\ref{eq:ccqe-q2qe}, whereas the technique relates
differential cross sections in \qq. Appendix~\ref{appendix:Q2QE}
shows the relationship between these two in the FG model.  The
differences are small compared to \qq bin widths for all relevant
kinematics in the \minerva experiment; however, in \mb, the
smearing becomes comparable to the bin width for $Q^2>0.2$~GeV$^2$.

The measurement of nuclear effects on carbon is extracted from the
public data releases for \minerva~\cite{minerva-nu-ccqe,
  minerva-antinu-ccqe}\footnote{The results used here correspond to
  the results with a flux estimate~\cite{new-minerva-flux} updated from the original
  publication, which predicts a significantly smaller flux and a
  smaller fractional flux uncertainty} and \mb~\cite{mbCCQE,
  mbAntiCCQE} using Equation~\ref{eq:ts} with standard propagation of
error techniques. For \minerva, the full covariance matrix, including cross-correlations, of
the
neutrino and antineutrino datasets is provided. For \mb, only the diagonals from the shape covariance matrices, and overall normalization factors are
provided separately for the neutrino and antineutrino datasets (which 
we assume to to be uncorrelated in this analysis). The data points and covariance matrices extracted in this work for both \minerva and \mb are available in the supplementary material.

\begin{figure}[h!]
  \centering
  \begin{subfigure}{0.45\textwidth}
    \includegraphics[width=\textwidth]{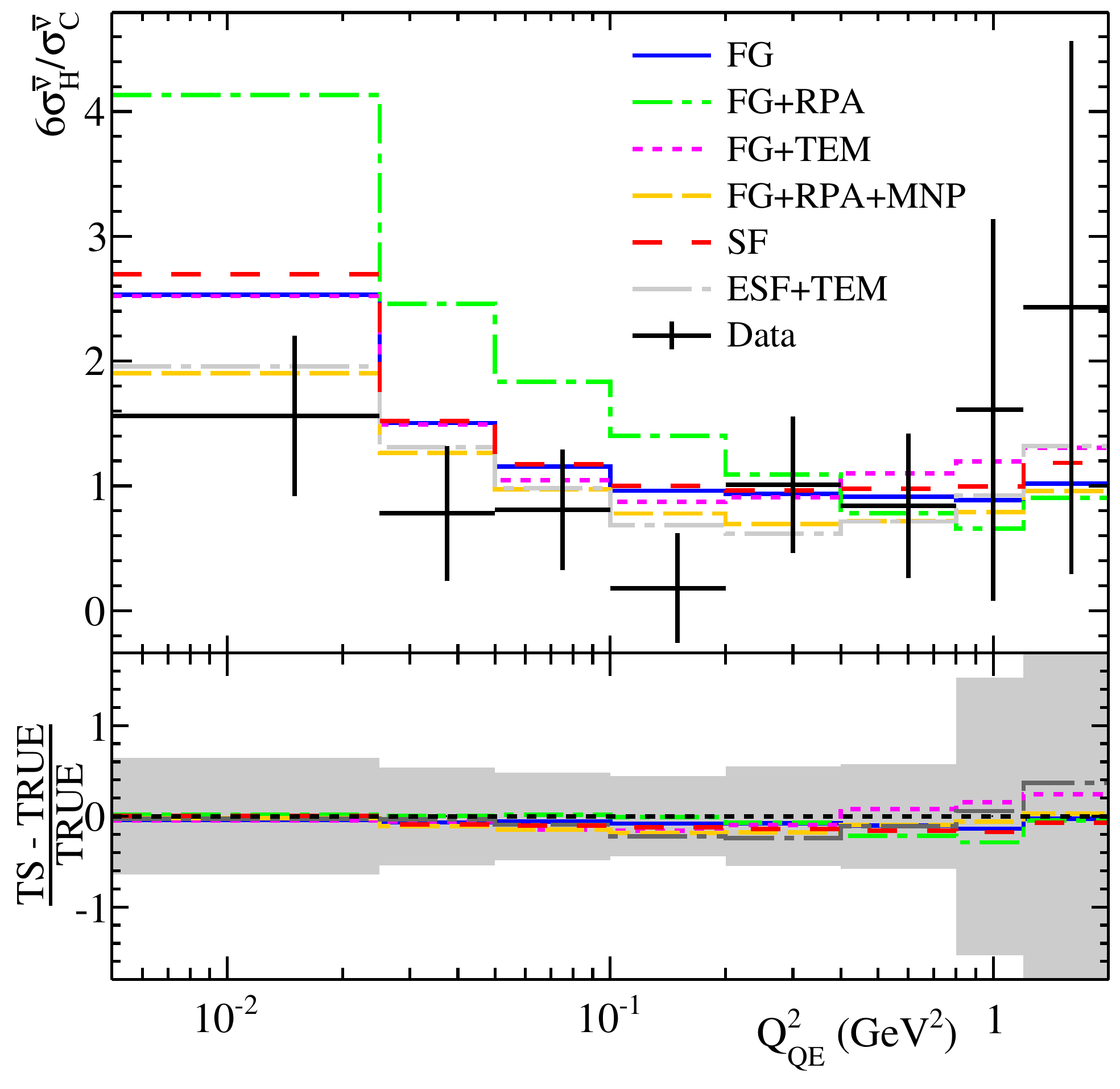}
    \caption{\minerva}
  \end{subfigure}
  \begin{subfigure}{0.45\textwidth}
    \includegraphics[width=\textwidth]{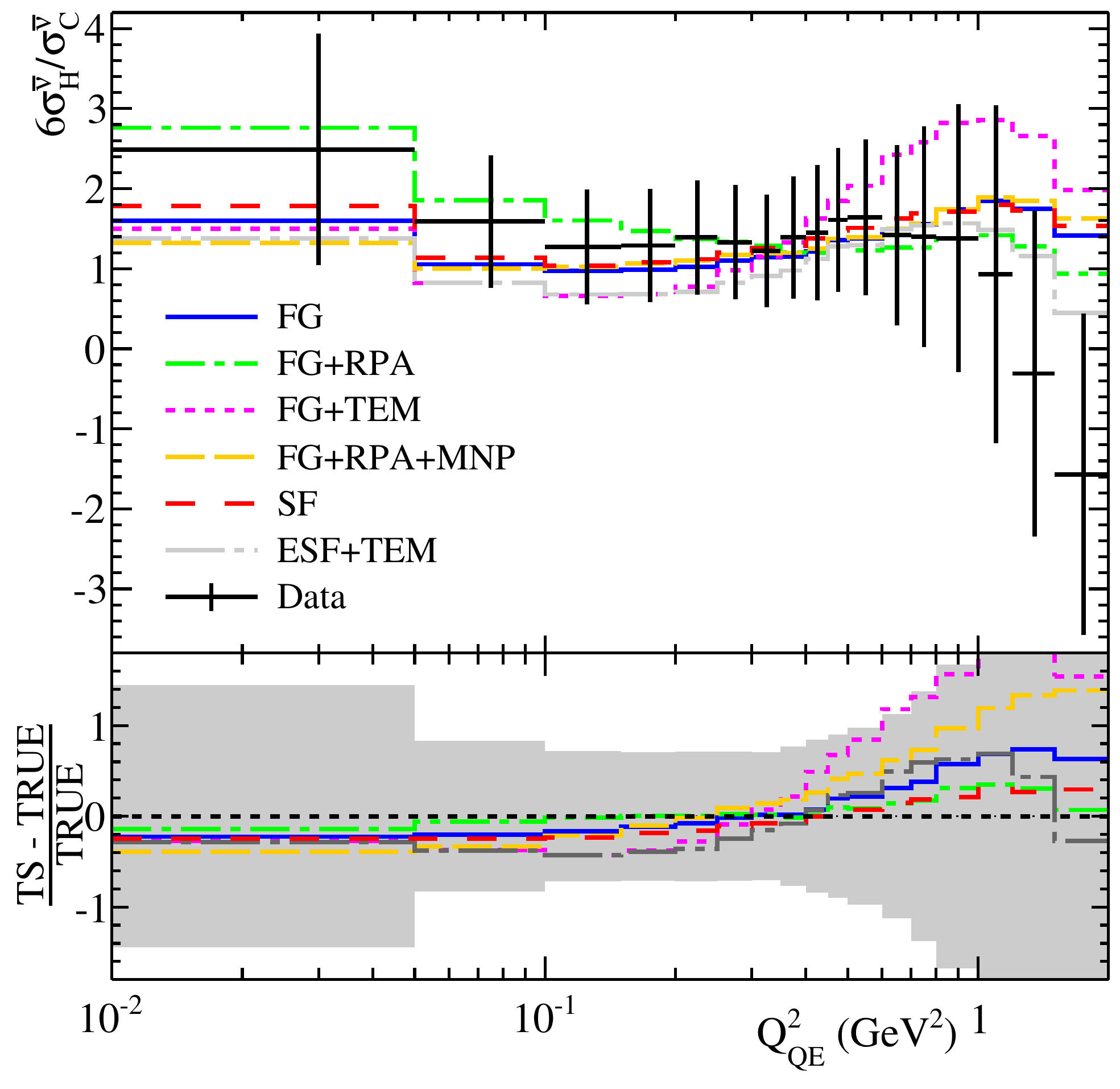}
    \caption{\mb}
    \end{subfigure}
  \caption{The value of $6\sigma^{\bar{\nu}}_{\mathrm{H}}/\sigma_{\mathrm{C}}^{\bar{\nu}}$ calculated using Equation~\ref{eq:ts} is shown for a variety of NEUT models, as well as for the extracted \minerva and \mb data. The model dependent bias on $6\sigma^{\bar{\nu}}_{\mathrm{H}}/\sigma_{\mathrm{C}}^{\bar{\nu}}$ is quantified by comparing the value obtained with Equation~\ref{eq:ts} (TS) with the exact value calculated for each model (TRUE). The bias, $\frac{\mathrm{TS} - \mathrm{TRUE}}{\mathrm{TRUE}}$, is compared with the fractional uncertainty on the measurement from data.}\label{fig:model_comparisons}
\end{figure}

In Figure~\ref{fig:model_comparisons}, the test statistic of Equation~\ref{eq:ts} is calculated for the \minerva and \mb data, and compared with the nuclear enhancement or suppression predicted by a variety of CCQE cross section models available in NEUT. The power of our measurement to constrain the choice of nuclear model is shown by the difference between our extracted data points and the ratio predicted by the various models tested. A $\chi^{2}$ value can be calculated for each model
\begin{align}
  \chi^{2} = \left(\nu_{i}^{\mbox{\scriptsize{DATA}}} - \nu_{i}^{\mbox{\scriptsize{MC}}} \right)M^{-1}_{ij}\left(\nu_{j}^{\mbox{\scriptsize{DATA}}} - \nu_{j}^{\mbox{\scriptsize{MC}}} \right),
  \label{eq:chi2}
\end{align}
\noindent where the measurement of nuclear effects from data is given
by $\nu_{i}^{\mbox{\scriptsize{DATA}}}$, the covariance matrix between
the data points is $M_{ij}$ and the NEUT prediction for each model is
given with $\nu_{i}^{\mbox{\scriptsize{MC}}}$. The $\chi^{2}$ values
for each model are given for both \minerva and \mb in
Table~\ref{tab:nominal_chi2}.  The models to which we compare the data
span calculational approaches to nuclear models for CCQE in
the literature, but are not a complete set. Any other model can be compared to the measurements in
this work using information in Appendix~\ref{appendix:applying}.

\begin{table}[h!]
\centering
{\renewcommand{\arraystretch}{1.2}
\begin{tabular}{c|cc}
  \hline\hline
  \multirow{2}{*}{Model} & \multicolumn{2}{c}{$\chi^{2}$/DOF} \\
  & \minerva & \mb \\
  \hline
  FG & 14.8/8 & 6.0/17 \\
  FG+RPA & 44.3/8 & 6.0/17 \\
  FG+RPA+MNP & 13.6/8 & 6.8/17 \\
  FG+TEM & 13.4/8 & 23.4/17 \\
  SF & 15.9/8 & 6.1/17 \\
  ESF+TEM & 12.8/8 & 6.2/17 \\
  \hline\hline
\end{tabular}}
\caption{$\chi^{2}$ values obtained with Equation~\ref{eq:chi2} for the various cross section models shown in Figure~\ref{fig:model_comparisons}.}\label{tab:nominal_chi2}
\end{table}

Any model dependent bias in the test statistic due to the free nucleon correction factor $\lambda(\qq)$ (see Equation~\ref{eq:ts-r} and Figure~\ref{fig:r_ratios}) or $\qq \rightarrow \qqqe$ differences (see Appendix~\ref{appendix:Q2QE}) can be calculated for each NEUT model by comparing the predicted ratio $6\sigma^{\bar{\nu}}_{\mathrm{H}}(\qqqe)/\sigma_{\mathrm{C}}^{\bar{\nu}}(\qqqe)$ for each model (labeled TRUE), with the test statistic (TS) calculated using Equation~\ref{eq:ts}. A large deviation between the TS and TRUE values would indicate that Equation~\ref{eq:ts} breaks down for that model and cannot be meaningfully compared with that model. The bottom panels of Figure~\ref{fig:model_comparisons} shows that this deviation is small compared to fractional uncertainties on the data for \minerva, but is large for \mb. Because the size of the bias for \minerva is small, certainly $<$10\% of the error on the data even in the highest \qqqe bins, we conclude that our extracted measurement of the enhancement and suppression in the $6\sigma^{\bar{\nu}}_{\mathrm{H}}(\qqqe)/\sigma_{\mathrm{C}}^{\bar{\nu}}(\qqqe)$ ratio can be used to differentiate between nuclear models.

\begin{figure}[htbp]
  \centering
    \includegraphics[width=0.45\textwidth]{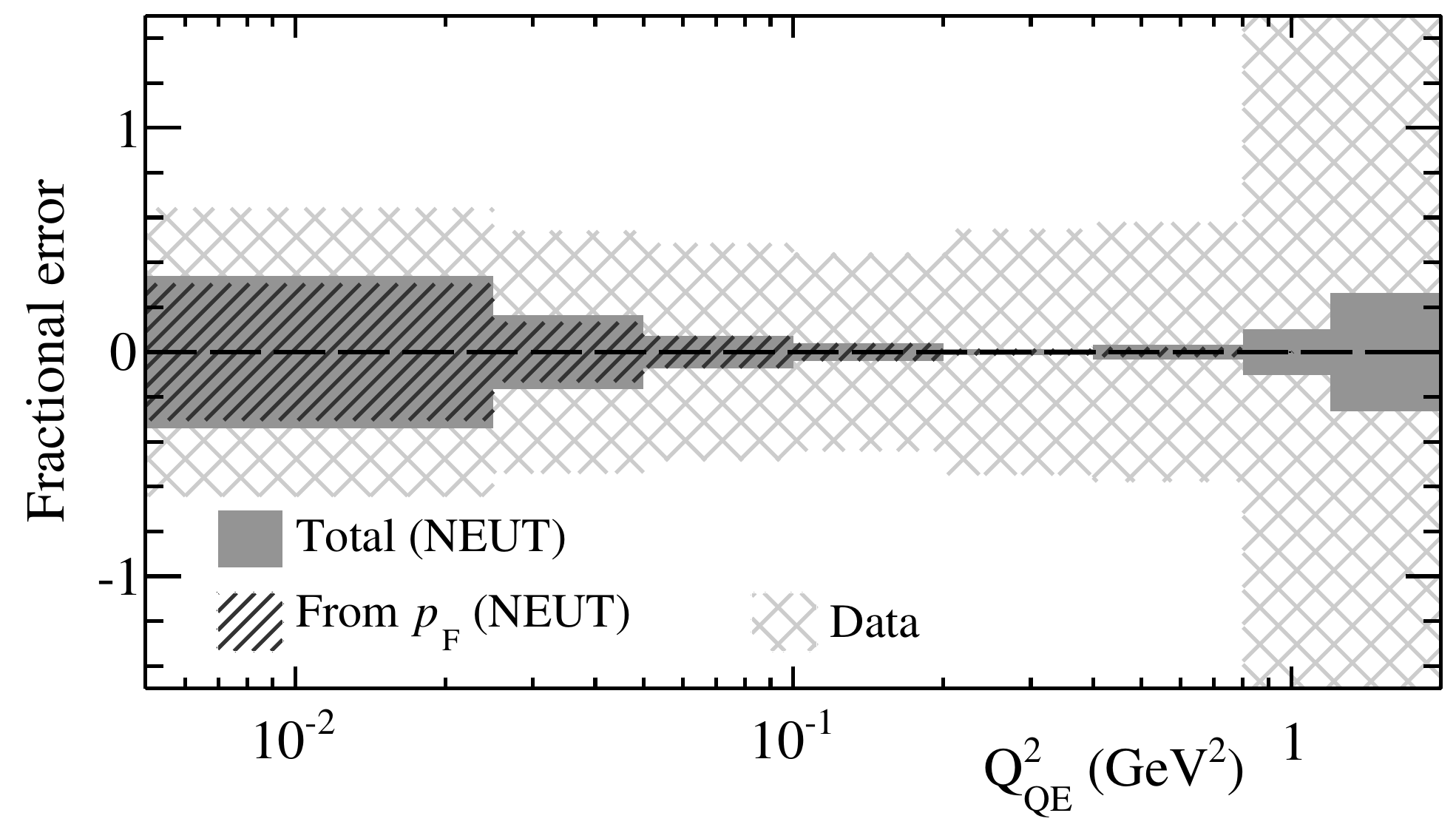}
    \caption{The fractional uncertainty on the value of $6\sigma^{\bar{\nu}}_{\mathrm{H}}/\sigma_{\mathrm{C}}^{\bar{\nu}}$ calculated for the FG model with \minerva. The total uncertainty is obtained by combining the 1$\sigma$ uncertainties in quadrature, and the dominant uncertainty, \pf is also shown separately. The fractional uncertainty on the data is shown for comparison.}\label{fig:RFG_model_error_ratio}
\end{figure}

Figure~\ref{fig:model_comparisons} and Table~\ref{tab:nominal_chi2} show that the extracted \minerva data have some power to differentiate between nuclear models, and that there is considerable tension between the data and all models tested. However, we have treated the NEUT nuclear models as having no free parameters, and have
calculated $\chi^{2}$ values assuming nominal model parameters. This tension may well be reduced by considering changes to the model parameters, and indeed this measurement could be used to tune the parameters of any one model.
Many of the models have no well defined theoretical uncertainties which can
be varied in NEUT; however, the FG model does have a number of
parameters which may be varied to estimate uncertainties within the base FG
model, and we may additionally consider uncertainties in the axial
form factor. To illustrate the possible reduction in tension due to modified nuclear model parameters, we consider variations in the FG of \ma = $1.00\pm0.02$ GeV~\cite{kuzmin_2006,bbba07,kuzmin_2008,Bodek:2007vi}, 
\pf = $217\pm5$ MeV~\cite{moniz_eA_RFG_1971}, \eb = $25\pm3$
MeV~\cite{moniz_eA_RFG_1971} and variations of $3$~MeV in \eb for {\it either}
neutrino or antineutrino to reflect uncertainty on whether the binding
energy is the same for neutrons and protons. Additionally, we consider the 3\% uncertainty on $F_{\mathrm{P}}(0)$ recommended in Reference~\cite{day_nue_numu_2012}; and take the difference between the non-dipole $F_{\mathrm{A}}$ from Reference~\cite{bbba07} and the dipole $F_{\mathrm{A}}$ as a 1$\sigma$ uncertainty. The uncertainties are combined in quadrature and compared to the fractional uncertainty on the data in Figure~\ref{fig:RFG_model_error_ratio}.
The FG model uncertainty is most significant at low \qq and is dominated by
the uncertainty on the Fermi momentum, \pf. 
As the model bias of our measurement is smallest at low \qq, changing \pf may improve the $\chi^{2}$ between our measurement and the predictions of the various FG based models considered in this work. 
We extend the $\chi^{2}$ calculation from Equation~\ref{eq:chi2} to
include a variable \pf parameter with a penalty term based on the \pf
uncertainty from electron-scattering data~\cite{moniz_eA_RFG_1971}. The best fit $\chi^{2}$ and \pf result
for each of the FG based models is shown in
Table~\ref{tab:fitted_chi2} for \minerva. The fit
reduces \pf slightly in order to reduce the value of
$6\sigma^{\bar{\nu}}_{\mathrm{H}}/\sigma^{\bar{\nu}}_{\mathrm{C}}$ at
low \qqqe, but there is no significant improvement in fit quality.
As already commented, this study is illustrative only, modifying nuclear model uncertainties may well significantly reduce the tension for other models, but it is interesting that in the case of simple FG-based nuclear models, the tensions cannot be significantly reduced by playing with the model uncertainties.

\begin{table}[t]
{\renewcommand{\arraystretch}{1.2}
\begin{tabular}{c|ccc}
  \hline\hline
  \multirow{2}{*}{Model} & \multicolumn{2}{c}{$\chi^{2}$/DOF} & \multirow{2}{*}{~~\pf (GeV$^{2}$)~~} \\
  & Nominal & ~~Fit~~ & \\
  \hline
  FG & 14.8 & 14.1 & $213.8 \pm 4.0$ \\
  FG+RPA & 44.3 & 38.2 & $207.6 \pm 4.0$ \\
  FG+RPA+MNP & 13.6 & 13.5 & $214.1 \pm 3.9$ \\
  FG+TEM & 13.4 & 12.8 & $215.8 \pm 4.5$ \\
  \hline\hline
\end{tabular}}
\caption{Best fit $\chi^{2}$ and \pf results for the fit to FG based models for \minerva data. The nominal $\chi^{2}$ with \pf = 217 MeV is included for comparison.
}\label{tab:fitted_chi2}
\end{table}

Improving the understanding of nuclear effects in neutrino scattering has become a focus for reducing systematic uncertainties in current and future neutrino oscillation experiments. As there are no current or future experiments which will take neutrino--nucleon scattering data in the few-GeV energy region, the method described here offers a unique opportunity to directly inspect the suppression or enhancement due to nuclear effects.  The method exploits the fact that antineutrinos have additional interactions on free protons (from the hydrogen), and corrects for neutrino and antineutrino flux and cross section differences. It was expected to work well at low \qq, and be relatively free of axial form factor or other uncertainties, and proves to be relatively unbiased at \minerva even at high \qq. Model dependent biases were seen for \mb, which should be borne in mind when applying this technique to other low energy datasets. The extracted measurement of nuclear effects in carbon is the first of its kind, and is easy to interpret for model builders. We conclude that models with nuclear screening due to long-range correlations must be balanced by the addition of multinucleon hard scattering processes, and that the combination of both effects is weakly favored over Fermi Gas models that only include the mean field of the nucleus. We also note that all of the models tested show considerable tension with the \minerva data. Constraints from this measurement could be improved using future, higher statistics, \minerva \ccqe measurements. This method could be applied to cross section measurements in terms of different kinematic variables, although a high-\qq bias will remain.

\FloatBarrier
\begin{acknowledgements}
This work was supported by the United States Department of Energy
under Grant DE-SC0008475 and by the Swiss National Science Foundation
and SERI. CW is grateful to the University of Rochester for
hospitality while this work was being carried out.  We thank Geralyn
Zeller for useful discussions about this technique during its early
development, and in particular for information about \mb's
consideration of a similar analysis.  We thank the
developers of the NEUT generator for implementation of many alternate
nuclear models, and the T2K collaboration for supporting this
development.  We thank the \minerva collaboration for early release of
their data corrected for the improved flux simulation.
\end{acknowledgements}
\bibliographystyle{apsrev4-1}
\bibliography{C12_nuclear_effects}

\begin{thebibliography}{57}%
\makeatletter
\providecommand \@ifxundefined [1]{%
 \@ifx{#1\undefined}
}%
\providecommand \@ifnum [1]{%
 \ifnum #1\expandafter \@firstoftwo
 \else \expandafter \@secondoftwo
 \fi
}%
\providecommand \@ifx [1]{%
 \ifx #1\expandafter \@firstoftwo
 \else \expandafter \@secondoftwo
 \fi
}%
\providecommand \natexlab [1]{#1}%
\providecommand \enquote  [1]{``#1''}%
\providecommand \bibnamefont  [1]{#1}%
\providecommand \bibfnamefont [1]{#1}%
\providecommand \citenamefont [1]{#1}%
\providecommand \href@noop [0]{\@secondoftwo}%
\providecommand \href [0]{\begingroup \@sanitize@url \@href}%
\providecommand \@href[1]{\@@startlink{#1}\@@href}%
\providecommand \@@href[1]{\endgroup#1\@@endlink}%
\providecommand \@sanitize@url [0]{\catcode `\\12\catcode `\$12\catcode
  `\&12\catcode `\#12\catcode `\^12\catcode `\_12\catcode `\%12\relax}%
\providecommand \@@startlink[1]{}%
\providecommand \@@endlink[0]{}%
\providecommand \url  [0]{\begingroup\@sanitize@url \@url }%
\providecommand \@url [1]{\endgroup\@href {#1}{\urlprefix }}%
\providecommand \urlprefix  [0]{URL }%
\providecommand \Eprint [0]{\href }%
\providecommand \doibase [0]{http://dx.doi.org/}%
\providecommand \selectlanguage [0]{\@gobble}%
\providecommand \bibinfo  [0]{\@secondoftwo}%
\providecommand \bibfield  [0]{\@secondoftwo}%
\providecommand \translation [1]{[#1]}%
\providecommand \BibitemOpen [0]{}%
\providecommand \bibitemStop [0]{}%
\providecommand \bibitemNoStop [0]{.\EOS\space}%
\providecommand \EOS [0]{\spacefactor3000\relax}%
\providecommand \BibitemShut  [1]{\csname bibitem#1\endcsname}%
\let\auto@bib@innerbib\@empty
\bibitem [{\citenamefont {Adler}(1963)}]{adler63}%
  \BibitemOpen
  \bibfield  {author} {\bibinfo {author} {\bibfnamefont {S.~L.}\ \bibnamefont
  {Adler}},\ }\href {\doibase 10.1007/BF02828811} {\bibfield  {journal}
  {\bibinfo  {journal} {Il Nuovo Cimento (1955-1965)}\ }\textbf {\bibinfo
  {volume} {30}},\ \bibinfo {pages} {1020} (\bibinfo {year}
  {1963})}\BibitemShut {NoStop}%
\bibitem [{\citenamefont {Marshak}\ \emph {et~al.}(1969)\citenamefont
  {Marshak}, \citenamefont {Riazuddin},\ and\ \citenamefont
  {Ryan}}]{marshak69}%
  \BibitemOpen
  \bibfield  {author} {\bibinfo {author} {\bibfnamefont {R.~E.}\ \bibnamefont
  {Marshak}}, \bibinfo {author} {\bibnamefont {Riazuddin}}, \ and\ \bibinfo
  {author} {\bibfnamefont {C.~P.}\ \bibnamefont {Ryan}},\ }\href@noop {} {\emph
  {\bibinfo {title} {Theory of Weak Interactions in Particle Physics}}}\
  (\bibinfo  {publisher} {Wiley-Interscience},\ \bibinfo {year} {1969})\ pp.\
  \bibinfo {pages} {1--761}\BibitemShut {NoStop}%
\bibitem [{\citenamefont {Pais}(1971)}]{pais71}%
  \BibitemOpen
  \bibfield  {author} {\bibinfo {author} {\bibfnamefont {A.}~\bibnamefont
  {Pais}},\ }\href {\doibase 10.1016/0003-4916(71)90018-2} {\bibfield
  {journal} {\bibinfo  {journal} {Annals Phys.}\ }\textbf {\bibinfo {volume}
  {63}},\ \bibinfo {pages} {361} (\bibinfo {year} {1971})}\BibitemShut
  {NoStop}%
\bibitem [{\citenamefont {Llewellyn~Smith}(1972)}]{smith72}%
  \BibitemOpen
  \bibfield  {author} {\bibinfo {author} {\bibfnamefont {C.}~\bibnamefont
  {Llewellyn~Smith}},\ }\href {\doibase 10.1016/0370-1573(72)90010-5}
  {\bibfield  {journal} {\bibinfo  {journal} {Phys. Rept.}\ }\textbf {\bibinfo
  {volume} {3}},\ \bibinfo {pages} {261} (\bibinfo {year} {1972})}\BibitemShut
  {NoStop}%
\bibitem [{\citenamefont {Budd}\ \emph {et~al.}(2003)\citenamefont {Budd},
  \citenamefont {Bodek},\ and\ \citenamefont {Arrington}}]{bba03}%
  \BibitemOpen
  \bibfield  {author} {\bibinfo {author} {\bibfnamefont {H.~S.}\ \bibnamefont
  {Budd}}, \bibinfo {author} {\bibfnamefont {A.}~\bibnamefont {Bodek}}, \ and\
  \bibinfo {author} {\bibfnamefont {J.}~\bibnamefont {Arrington}},\ }in\
  \href@noop {} {\emph {\bibinfo {booktitle} {{2nd International Workshop on
  Neutrino-Nucleus Interactions in the Few GeV Region (NuInt 02) Irvine,
  California, December 12-15, 2002}}}}\ (\bibinfo {year} {2003})\ \Eprint
  {http://arxiv.org/abs/hep-ex/0308005} {arXiv:hep-ex/0308005 [hep-ex]}
  \BibitemShut {NoStop}%
\bibitem [{\citenamefont {Kuzmin}\ \emph {et~al.}(2006)\citenamefont {Kuzmin},
  \citenamefont {Lyubushkin},\ and\ \citenamefont {Naumov}}]{kuzmin_2006}%
  \BibitemOpen
  \bibfield  {author} {\bibinfo {author} {\bibfnamefont {K.~S.}\ \bibnamefont
  {Kuzmin}}, \bibinfo {author} {\bibfnamefont {V.~V.}\ \bibnamefont
  {Lyubushkin}}, \ and\ \bibinfo {author} {\bibfnamefont {V.~A.}\ \bibnamefont
  {Naumov}},\ }\href@noop {} {\bibfield  {journal} {\bibinfo  {journal} {Acta
  Phys. Polon.}\ }\textbf {\bibinfo {volume} {B37}},\ \bibinfo {pages} {2337}
  (\bibinfo {year} {2006})},\ \Eprint {http://arxiv.org/abs/hep-ph/0606184}
  {arXiv:hep-ph/0606184 [hep-ph]} \BibitemShut {NoStop}%
\bibitem [{\citenamefont {Bodek}\ \emph {et~al.}()\citenamefont {Bodek},
  \citenamefont {Avvakumov}, \citenamefont {Bradford},\ and\ \citenamefont
  {Budd}}]{bbba07}%
  \BibitemOpen
  \bibfield  {author} {\bibinfo {author} {\bibfnamefont {A.}~\bibnamefont
  {Bodek}}, \bibinfo {author} {\bibfnamefont {S.}~\bibnamefont {Avvakumov}},
  \bibinfo {author} {\bibfnamefont {R.}~\bibnamefont {Bradford}}, \ and\
  \bibinfo {author} {\bibfnamefont {H.}~\bibnamefont {Budd}},\ }\href {\doibase
  10.1140/epjc/s10052-007-0491-4} {\bibfield  {journal} {\bibinfo  {journal}
  {The European Physical Journal C}\ }\textbf {\bibinfo {volume} {53}},\
  \bibinfo {pages} {349}}\BibitemShut {NoStop}%
\bibitem [{\citenamefont {Kuzmin}\ \emph {et~al.}(2008)\citenamefont {Kuzmin},
  \citenamefont {Lyubushkin},\ and\ \citenamefont {Naumov}}]{kuzmin_2008}%
  \BibitemOpen
  \bibfield  {author} {\bibinfo {author} {\bibfnamefont {K.~S.}\ \bibnamefont
  {Kuzmin}}, \bibinfo {author} {\bibfnamefont {V.~V.}\ \bibnamefont
  {Lyubushkin}}, \ and\ \bibinfo {author} {\bibfnamefont {V.~A.}\ \bibnamefont
  {Naumov}},\ }\href {\doibase 10.1140/epjc/s10052-008-0582-x} {\bibfield
  {journal} {\bibinfo  {journal} {Eur. Phys. J.}\ }\textbf {\bibinfo {volume}
  {C54}},\ \bibinfo {pages} {517} (\bibinfo {year} {2008})},\ \Eprint
  {http://arxiv.org/abs/0712.4384} {arXiv:0712.4384 [hep-ph]} \BibitemShut
  {NoStop}%
\bibitem [{\citenamefont {Bodek}\ \emph {et~al.}(2008)\citenamefont {Bodek},
  \citenamefont {Avvakumov}, \citenamefont {Bradford},\ and\ \citenamefont
  {Budd}}]{Bodek:2007vi}%
  \BibitemOpen
  \bibfield  {author} {\bibinfo {author} {\bibfnamefont {A.}~\bibnamefont
  {Bodek}}, \bibinfo {author} {\bibfnamefont {S.}~\bibnamefont {Avvakumov}},
  \bibinfo {author} {\bibfnamefont {R.}~\bibnamefont {Bradford}}, \ and\
  \bibinfo {author} {\bibfnamefont {H.~S.}\ \bibnamefont {Budd}},\ }\href
  {\doibase 10.1088/1742-6596/110/8/082004} {\bibfield  {journal} {\bibinfo
  {journal} {J. Phys. Conf. Ser.}\ }\textbf {\bibinfo {volume} {110}},\
  \bibinfo {pages} {082004} (\bibinfo {year} {2008})},\ \Eprint
  {http://arxiv.org/abs/0709.3538} {arXiv:0709.3538 [hep-ex]} \BibitemShut
  {NoStop}%
\bibitem [{\citenamefont {Smith}\ and\ \citenamefont
  {Moniz}(1972)}]{smith-moniz72}%
  \BibitemOpen
  \bibfield  {author} {\bibinfo {author} {\bibfnamefont {R.}~\bibnamefont
  {Smith}}\ and\ \bibinfo {author} {\bibfnamefont {E.}~\bibnamefont {Moniz}},\
  }\href {\doibase 10.1016/0550-3213(72)90040-5} {\bibfield  {journal}
  {\bibinfo  {journal} {Nucl. Phys.}\ }\textbf {\bibinfo {volume} {B43}},\
  \bibinfo {pages} {605} (\bibinfo {year} {1972})}\BibitemShut {NoStop}%
\bibitem [{\citenamefont {Aguilar-Arevalo}\ \emph {et~al.}(2010)\citenamefont
  {Aguilar-Arevalo} \emph {et~al.}}]{mbCCQE}%
  \BibitemOpen
  \bibfield  {author} {\bibinfo {author} {\bibfnamefont {A.}~\bibnamefont
  {Aguilar-Arevalo}} \emph {et~al.} (\bibinfo {collaboration} {MiniBooNE
  Collaboration}),\ }\href {\doibase 10.1103/PhysRevD.81.092005} {\bibfield
  {journal} {\bibinfo  {journal} {Phys. Rev.}\ }\textbf {\bibinfo {volume}
  {D81}},\ \bibinfo {pages} {092005} (\bibinfo {year} {2010})},\ \Eprint
  {http://arxiv.org/abs/1002.2680} {arXiv:1002.2680 [hep-ex]} \BibitemShut
  {NoStop}%
\bibitem [{\citenamefont {Gran}\ \emph {et~al.}(2006)\citenamefont {Gran} \emph
  {et~al.}}]{k2kCCQE}%
  \BibitemOpen
  \bibfield  {author} {\bibinfo {author} {\bibfnamefont {R.}~\bibnamefont
  {Gran}} \emph {et~al.} (\bibinfo {collaboration} {K2K}),\ }\href {\doibase
  10.1103/PhysRevD.74.052002} {\bibfield  {journal} {\bibinfo  {journal} {Phys.
  Rev.}\ }\textbf {\bibinfo {volume} {D74}},\ \bibinfo {pages} {052002}
  (\bibinfo {year} {2006})},\ \Eprint {http://arxiv.org/abs/hep-ex/0603034}
  {arXiv:hep-ex/0603034 [hep-ex]} \BibitemShut {NoStop}%
\bibitem [{\citenamefont {Lyubushkin}\ \emph {et~al.}(2009)\citenamefont
  {Lyubushkin} \emph {et~al.}}]{NOMAD_CCQE}%
  \BibitemOpen
  \bibfield  {author} {\bibinfo {author} {\bibfnamefont {V.}~\bibnamefont
  {Lyubushkin}} \emph {et~al.},\ }\href {\doibase
  10.1140/epjc/s10052-009-1113-0} {\bibfield  {journal} {\bibinfo  {journal}
  {The European Physical Journal C}\ }\textbf {\bibinfo {volume} {63}},\
  \bibinfo {pages} {355} (\bibinfo {year} {2009})}\BibitemShut {NoStop}%
\bibitem [{\citenamefont {Dorman}(2009)}]{minosCCQE}%
  \BibitemOpen
  \bibfield  {author} {\bibinfo {author} {\bibfnamefont {M.}~\bibnamefont
  {Dorman}} (\bibinfo {collaboration} {MINOS Collaboration}),\ }\href {\doibase
  10.1063/1.3274143} {\bibfield  {journal} {\bibinfo  {journal} {AIP Conf.
  Proc.}\ }\textbf {\bibinfo {volume} {1189}},\ \bibinfo {pages} {133}
  (\bibinfo {year} {2009})}\BibitemShut {NoStop}%
\bibitem [{\citenamefont {Fiorentini}\ \emph {et~al.}(2013)\citenamefont
  {Fiorentini} \emph {et~al.}}]{minerva-nu-ccqe}%
  \BibitemOpen
  \bibfield  {author} {\bibinfo {author} {\bibfnamefont {G.}~\bibnamefont
  {Fiorentini}} \emph {et~al.} (\bibinfo {collaboration} {{MINER$\nu$A}
  Collaboration}),\ }\href {\doibase 10.1103/PhysRevLett.111.022502} {\bibfield
   {journal} {\bibinfo  {journal} {Phys. Rev. Lett.}\ }\textbf {\bibinfo
  {volume} {111}},\ \bibinfo {pages} {022502} (\bibinfo {year} {2013})},\
  \Eprint {http://arxiv.org/abs/1305.2243} {arXiv:1305.2243 [hep-ex]}
  \BibitemShut {NoStop}%
\bibitem [{\citenamefont {Fields}\ \emph {et~al.}(2013)\citenamefont {Fields}
  \emph {et~al.}}]{minerva-antinu-ccqe}%
  \BibitemOpen
  \bibfield  {author} {\bibinfo {author} {\bibfnamefont {L.}~\bibnamefont
  {Fields}} \emph {et~al.} (\bibinfo {collaboration} {{MINER$\nu$A}
  Collaboration}),\ }\href {\doibase 10.1103/PhysRevLett.111.022501} {\bibfield
   {journal} {\bibinfo  {journal} {Phys. Rev. Lett.}\ }\textbf {\bibinfo
  {volume} {111}},\ \bibinfo {pages} {022501} (\bibinfo {year} {2013})},\
  \Eprint {http://arxiv.org/abs/1305.2234} {arXiv:1305.2234 [hep-ex]}
  \BibitemShut {NoStop}%
\bibitem [{\citenamefont {Aguilar-Arevalo}\ \emph {et~al.}(2013)\citenamefont
  {Aguilar-Arevalo} \emph {et~al.}}]{mbAntiCCQE}%
  \BibitemOpen
  \bibfield  {author} {\bibinfo {author} {\bibfnamefont {A.}~\bibnamefont
  {Aguilar-Arevalo}} \emph {et~al.} (\bibinfo {collaboration} {MiniBooNE
  Collaboration}),\ }\href {\doibase 10.1103/PhysRevD.88.032001} {\bibfield
  {journal} {\bibinfo  {journal} {Phys. Rev.}\ }\textbf {\bibinfo {volume}
  {D88}},\ \bibinfo {pages} {032001} (\bibinfo {year} {2013})},\ \Eprint
  {http://arxiv.org/abs/1301.7067} {arXiv:1301.7067 [hep-ex]} \BibitemShut
  {NoStop}%
\bibitem [{\citenamefont {Abe}\ \emph {et~al.}(2015{\natexlab{a}})\citenamefont
  {Abe} \emph {et~al.}}]{ingridCCQE}%
  \BibitemOpen
  \bibfield  {author} {\bibinfo {author} {\bibfnamefont {K.}~\bibnamefont
  {Abe}} \emph {et~al.} (\bibinfo {collaboration} {T2K}),\ }\href {\doibase
  10.1103/PhysRevD.91.112002} {\bibfield  {journal} {\bibinfo  {journal} {Phys.
  Rev.}\ }\textbf {\bibinfo {volume} {D91}},\ \bibinfo {pages} {112002}
  (\bibinfo {year} {2015}{\natexlab{a}})},\ \Eprint
  {http://arxiv.org/abs/1503.07452} {arXiv:1503.07452 [hep-ex]} \BibitemShut
  {NoStop}%
\bibitem [{\citenamefont {Abe}\ \emph {et~al.}(2015{\natexlab{b}})\citenamefont
  {Abe} \emph {et~al.}}]{t2kCCQE}%
  \BibitemOpen
  \bibfield  {author} {\bibinfo {author} {\bibfnamefont {K.}~\bibnamefont
  {Abe}} \emph {et~al.} (\bibinfo {collaboration} {T2K}),\ }\href {\doibase
  10.1103/PhysRevD.92.112003} {\bibfield  {journal} {\bibinfo  {journal} {Phys.
  Rev.}\ }\textbf {\bibinfo {volume} {D92}},\ \bibinfo {pages} {112003}
  (\bibinfo {year} {2015}{\natexlab{b}})},\ \Eprint
  {http://arxiv.org/abs/1411.6264} {arXiv:1411.6264 [hep-ex]} \BibitemShut
  {NoStop}%
\bibitem [{\citenamefont {Rodrigues}\ \emph {et~al.}(2016)\citenamefont
  {Rodrigues} \emph {et~al.}}]{Rodrigues:2015hik}%
  \BibitemOpen
  \bibfield  {author} {\bibinfo {author} {\bibfnamefont {P.~A.}\ \bibnamefont
  {Rodrigues}} \emph {et~al.} (\bibinfo {collaboration} {MINERvA}),\ }\href
  {\doibase 10.1103/PhysRevLett.116.071802} {\bibfield  {journal} {\bibinfo
  {journal} {Phys. Rev. Lett.}\ }\textbf {\bibinfo {volume} {116}},\ \bibinfo
  {pages} {071802} (\bibinfo {year} {2016})},\ \Eprint
  {http://arxiv.org/abs/1511.05944} {arXiv:1511.05944 [hep-ex]} \BibitemShut
  {NoStop}%
\bibitem [{\citenamefont {Tice}\ \emph {et~al.}(2014)\citenamefont {Tice} \emph
  {et~al.}}]{minerva_ccinc_2014}%
  \BibitemOpen
  \bibfield  {author} {\bibinfo {author} {\bibfnamefont {B.}~\bibnamefont
  {Tice}} \emph {et~al.} (\bibinfo {collaboration} {{MINER$\nu$A
  Collaboration}}),\ }\href {\doibase 10.1103/PhysRevLett.112.231801}
  {\bibfield  {journal} {\bibinfo  {journal} {Phys. Rev. Lett.}\ }\textbf
  {\bibinfo {volume} {112}},\ \bibinfo {pages} {231801} (\bibinfo {year}
  {2014})},\ \Eprint {http://arxiv.org/abs/1403.2103} {arXiv:1403.2103
  [hep-ex]} \BibitemShut {NoStop}%
\bibitem [{\citenamefont {Benhar}\ and\ \citenamefont {Fabrocini}(2000)}]{sf}%
  \BibitemOpen
  \bibfield  {author} {\bibinfo {author} {\bibfnamefont {O.}~\bibnamefont
  {Benhar}}\ and\ \bibinfo {author} {\bibfnamefont {A.}~\bibnamefont
  {Fabrocini}},\ }\href {\doibase 10.1103/PhysRevC.62.034304} {\bibfield
  {journal} {\bibinfo  {journal} {Phys. Rev.}\ }\textbf {\bibinfo {volume}
  {C62}},\ \bibinfo {pages} {034304} (\bibinfo {year} {2000})},\ \Eprint
  {http://arxiv.org/abs/nucl-th/9909014} {arXiv:nucl-th/9909014 [nucl-th]}
  \BibitemShut {NoStop}%
\bibitem [{\citenamefont {Ankowski}\ and\ \citenamefont
  {Sobczyk}(2006)}]{ankowski_SF}%
  \BibitemOpen
  \bibfield  {author} {\bibinfo {author} {\bibfnamefont {A.~M.}\ \bibnamefont
  {Ankowski}}\ and\ \bibinfo {author} {\bibfnamefont {J.~T.}\ \bibnamefont
  {Sobczyk}},\ }\href {\doibase 10.1103/PhysRevC.74.054316} {\bibfield
  {journal} {\bibinfo  {journal} {Phys. Rev. C}\ }\textbf {\bibinfo {volume}
  {74}},\ \bibinfo {pages} {054316} (\bibinfo {year} {2006})}\BibitemShut
  {NoStop}%
\bibitem [{\citenamefont {Butkevich}(2009)}]{butkevich_2009}%
  \BibitemOpen
  \bibfield  {author} {\bibinfo {author} {\bibfnamefont {A.}~\bibnamefont
  {Butkevich}},\ }\href {\doibase 10.1103/PhysRevC.80.014610} {\bibfield
  {journal} {\bibinfo  {journal} {Phys. Rev.}\ }\textbf {\bibinfo {volume}
  {C80}},\ \bibinfo {pages} {014610} (\bibinfo {year} {2009})},\ \Eprint
  {http://arxiv.org/abs/0904.1472} {arXiv:0904.1472 [nucl-th]} \BibitemShut
  {NoStop}%
\bibitem [{\citenamefont {Bodek}\ \emph {et~al.}(2014)\citenamefont {Bodek},
  \citenamefont {Christy},\ and\ \citenamefont {Coopersmith}}]{eff-sf}%
  \BibitemOpen
  \bibfield  {author} {\bibinfo {author} {\bibfnamefont {A.}~\bibnamefont
  {Bodek}}, \bibinfo {author} {\bibfnamefont {M.~E.}\ \bibnamefont {Christy}},
  \ and\ \bibinfo {author} {\bibfnamefont {B.}~\bibnamefont {Coopersmith}},\
  }\href {\doibase 10.1140/epjc/s10052-014-3091-0} {\bibfield  {journal}
  {\bibinfo  {journal} {Eur. Phys. J.}\ }\textbf {\bibinfo {volume} {C74}},\
  \bibinfo {pages} {3091} (\bibinfo {year} {2014})},\ \Eprint
  {http://arxiv.org/abs/1405.0583} {arXiv:1405.0583 [hep-ph]} \BibitemShut
  {NoStop}%
\bibitem [{\citenamefont {Leitner}\ \emph {et~al.}(2009)\citenamefont
  {Leitner}, \citenamefont {Buss}, \citenamefont {Alvarez-Ruso},\ and\
  \citenamefont {Mosel}}]{leitner_2009}%
  \BibitemOpen
  \bibfield  {author} {\bibinfo {author} {\bibfnamefont {T.}~\bibnamefont
  {Leitner}}, \bibinfo {author} {\bibfnamefont {O.}~\bibnamefont {Buss}},
  \bibinfo {author} {\bibfnamefont {L.}~\bibnamefont {Alvarez-Ruso}}, \ and\
  \bibinfo {author} {\bibfnamefont {U.}~\bibnamefont {Mosel}},\ }\href
  {\doibase 10.1103/PhysRevC.79.034601} {\bibfield  {journal} {\bibinfo
  {journal} {Phys. Rev. C}\ }\textbf {\bibinfo {volume} {79}},\ \bibinfo
  {pages} {034601} (\bibinfo {year} {2009})}\BibitemShut {NoStop}%
\bibitem [{\citenamefont {Maieron}\ \emph {et~al.}(2003)\citenamefont
  {Maieron}, \citenamefont {Martinez}, \citenamefont {Caballero},\ and\
  \citenamefont {Udias}}]{madrid_2003}%
  \BibitemOpen
  \bibfield  {author} {\bibinfo {author} {\bibfnamefont {C.}~\bibnamefont
  {Maieron}}, \bibinfo {author} {\bibfnamefont {M.}~\bibnamefont {Martinez}},
  \bibinfo {author} {\bibfnamefont {J.}~\bibnamefont {Caballero}}, \ and\
  \bibinfo {author} {\bibfnamefont {J.}~\bibnamefont {Udias}},\ }\href
  {\doibase 10.1103/PhysRevC.68.048501} {\bibfield  {journal} {\bibinfo
  {journal} {Phys. Rev.}\ }\textbf {\bibinfo {volume} {C68}},\ \bibinfo {pages}
  {048501} (\bibinfo {year} {2003})},\ \Eprint
  {http://arxiv.org/abs/nucl-th/0303075} {arXiv:nucl-th/0303075 [nucl-th]}
  \BibitemShut {NoStop}%
\bibitem [{\citenamefont {Meucci}\ \emph {et~al.}(2004)\citenamefont {Meucci},
  \citenamefont {Giusti},\ and\ \citenamefont {Pacati}}]{meucci_2003}%
  \BibitemOpen
  \bibfield  {author} {\bibinfo {author} {\bibfnamefont {A.}~\bibnamefont
  {Meucci}}, \bibinfo {author} {\bibfnamefont {C.}~\bibnamefont {Giusti}}, \
  and\ \bibinfo {author} {\bibfnamefont {F.~D.}\ \bibnamefont {Pacati}},\
  }\href {\doibase 10.1016/j.nuclphysa.2004.04.108} {\bibfield  {journal}
  {\bibinfo  {journal} {Nucl. Phys.}\ }\textbf {\bibinfo {volume} {A739}},\
  \bibinfo {pages} {277} (\bibinfo {year} {2004})},\ \Eprint
  {http://arxiv.org/abs/nucl-th/0311081} {arXiv:nucl-th/0311081 [nucl-th]}
  \BibitemShut {NoStop}%
\bibitem [{\citenamefont {Pandey}\ \emph {et~al.}(2015)\citenamefont {Pandey},
  \citenamefont {Jachowicz}, \citenamefont {Van~Cuyck}, \citenamefont
  {Ryckebusch},\ and\ \citenamefont {Martini}}]{pandey_2014}%
  \BibitemOpen
  \bibfield  {author} {\bibinfo {author} {\bibfnamefont {V.}~\bibnamefont
  {Pandey}}, \bibinfo {author} {\bibfnamefont {N.}~\bibnamefont {Jachowicz}},
  \bibinfo {author} {\bibfnamefont {T.}~\bibnamefont {Van~Cuyck}}, \bibinfo
  {author} {\bibfnamefont {J.}~\bibnamefont {Ryckebusch}}, \ and\ \bibinfo
  {author} {\bibfnamefont {M.}~\bibnamefont {Martini}},\ }\href {\doibase
  10.1103/PhysRevC.92.024606} {\bibfield  {journal} {\bibinfo  {journal} {Phys.
  Rev.}\ }\textbf {\bibinfo {volume} {C92}},\ \bibinfo {pages} {024606}
  (\bibinfo {year} {2015})},\ \Eprint {http://arxiv.org/abs/1412.4624}
  {arXiv:1412.4624 [nucl-th]} \BibitemShut {NoStop}%
\bibitem [{\citenamefont {Nieves}\ \emph {et~al.}(2011)\citenamefont {Nieves},
  \citenamefont {Simo},\ and\ \citenamefont {Vacas}}]{nieves_2011}%
  \BibitemOpen
  \bibfield  {author} {\bibinfo {author} {\bibfnamefont {J.}~\bibnamefont
  {Nieves}}, \bibinfo {author} {\bibfnamefont {I.~R.}\ \bibnamefont {Simo}}, \
  and\ \bibinfo {author} {\bibfnamefont {M.~J.~V.}\ \bibnamefont {Vacas}},\
  }\href {\doibase 10.1103/PhysRevC.83.045501} {\bibfield  {journal} {\bibinfo
  {journal} {Phys. Rev. C}\ }\textbf {\bibinfo {volume} {83}},\ \bibinfo
  {pages} {045501} (\bibinfo {year} {2011})}\BibitemShut {NoStop}%
\bibitem [{\citenamefont {Martini}\ \emph {et~al.}(2009)\citenamefont
  {Martini}, \citenamefont {Ericson}, \citenamefont {Chanfray},\ and\
  \citenamefont {Marteau}}]{martini_2009}%
  \BibitemOpen
  \bibfield  {author} {\bibinfo {author} {\bibfnamefont {M.}~\bibnamefont
  {Martini}}, \bibinfo {author} {\bibfnamefont {M.}~\bibnamefont {Ericson}},
  \bibinfo {author} {\bibfnamefont {G.}~\bibnamefont {Chanfray}}, \ and\
  \bibinfo {author} {\bibfnamefont {J.}~\bibnamefont {Marteau}},\ }\href
  {\doibase 10.1103/PhysRevC.80.065501} {\bibfield  {journal} {\bibinfo
  {journal} {Phys. Rev.}\ }\textbf {\bibinfo {volume} {C80}},\ \bibinfo {pages}
  {065501} (\bibinfo {year} {2009})},\ \Eprint {http://arxiv.org/abs/0910.2622}
  {arXiv:0910.2622 [nucl-th]} \BibitemShut {NoStop}%
\bibitem [{\citenamefont {Formaggio}\ and\ \citenamefont
  {Zeller}(2012)}]{zeller12}%
  \BibitemOpen
  \bibfield  {author} {\bibinfo {author} {\bibfnamefont {J.~A.}\ \bibnamefont
  {Formaggio}}\ and\ \bibinfo {author} {\bibfnamefont {G.~P.}\ \bibnamefont
  {Zeller}},\ }\href {\doibase 10.1103/RevModPhys.84.1307} {\bibfield
  {journal} {\bibinfo  {journal} {Rev. Mod. Phys.}\ }\textbf {\bibinfo {volume}
  {84}},\ \bibinfo {pages} {1307} (\bibinfo {year} {2012})}\BibitemShut
  {NoStop}%
\bibitem [{\citenamefont {Alvarez-Ruso}\ \emph {et~al.}(2014)\citenamefont
  {Alvarez-Ruso}, \citenamefont {Hayato},\ and\ \citenamefont
  {Nieves}}]{hayato_review_2014}%
  \BibitemOpen
  \bibfield  {author} {\bibinfo {author} {\bibfnamefont {L.}~\bibnamefont
  {Alvarez-Ruso}}, \bibinfo {author} {\bibfnamefont {Y.}~\bibnamefont
  {Hayato}}, \ and\ \bibinfo {author} {\bibfnamefont {J.}~\bibnamefont
  {Nieves}},\ }\href {\doibase 10.1088/1367-2630/16/7/075015} {\bibfield
  {journal} {\bibinfo  {journal} {New J. Phys.}\ }\textbf {\bibinfo {volume}
  {16}},\ \bibinfo {pages} {075015} (\bibinfo {year} {2014})},\ \Eprint
  {http://arxiv.org/abs/1403.2673} {arXiv:1403.2673 [hep-ph]} \BibitemShut
  {NoStop}%
\bibitem [{\citenamefont {Garvey}\ \emph {et~al.}(2015)\citenamefont {Garvey},
  \citenamefont {Harris}, \citenamefont {Tanaka}, \citenamefont {Tayloe},\ and\
  \citenamefont {Zeller}}]{garvey_review_2014}%
  \BibitemOpen
  \bibfield  {author} {\bibinfo {author} {\bibfnamefont {G.~T.}\ \bibnamefont
  {Garvey}}, \bibinfo {author} {\bibfnamefont {D.~A.}\ \bibnamefont {Harris}},
  \bibinfo {author} {\bibfnamefont {H.~A.}\ \bibnamefont {Tanaka}}, \bibinfo
  {author} {\bibfnamefont {R.}~\bibnamefont {Tayloe}}, \ and\ \bibinfo {author}
  {\bibfnamefont {G.~P.}\ \bibnamefont {Zeller}},\ }\href {\doibase
  10.1016/j.physrep.2015.04.001} {\bibfield  {journal} {\bibinfo  {journal}
  {Phys. Rept.}\ }\textbf {\bibinfo {volume} {580}},\ \bibinfo {pages} {1}
  (\bibinfo {year} {2015})},\ \Eprint {http://arxiv.org/abs/1412.4294}
  {arXiv:1412.4294 [hep-ex]} \BibitemShut {NoStop}%
\bibitem [{\citenamefont {Aguilar-Arevalo}\ \emph
  {et~al.}(2009{\natexlab{a}})\citenamefont {Aguilar-Arevalo} \emph
  {et~al.}}]{mb_nue_app_2009}%
  \BibitemOpen
  \bibfield  {author} {\bibinfo {author} {\bibfnamefont {A.}~\bibnamefont
  {Aguilar-Arevalo}} \emph {et~al.} (\bibinfo {collaboration} {MiniBooNE
  Collaboration}),\ }\href {\doibase 10.1103/PhysRevLett.102.101802} {\bibfield
   {journal} {\bibinfo  {journal} {Phys. Rev. Lett.}\ }\textbf {\bibinfo
  {volume} {102}},\ \bibinfo {pages} {101802} (\bibinfo {year}
  {2009}{\natexlab{a}})},\ \Eprint {http://arxiv.org/abs/0812.2243}
  {arXiv:0812.2243 [hep-ex]} \BibitemShut {NoStop}%
\bibitem [{\citenamefont {Abe}\ \emph {et~al.}(2011{\natexlab{a}})\citenamefont
  {Abe} \emph {et~al.}}]{t2k_nue_2011}%
  \BibitemOpen
  \bibfield  {author} {\bibinfo {author} {\bibfnamefont {K.}~\bibnamefont
  {Abe}} \emph {et~al.} (\bibinfo {collaboration} {T2K Collaboration}),\ }\href
  {\doibase 10.1103/PhysRevLett.107.041801} {\bibfield  {journal} {\bibinfo
  {journal} {Phys. Rev. Lett.}\ }\textbf {\bibinfo {volume} {107}},\ \bibinfo
  {pages} {041801} (\bibinfo {year} {2011}{\natexlab{a}})}\BibitemShut
  {NoStop}%
\bibitem [{\citenamefont {Ayres}\ \emph {et~al.}(2005)\citenamefont {Ayres}
  \emph {et~al.}}]{nova}%
  \BibitemOpen
  \bibfield  {author} {\bibinfo {author} {\bibfnamefont {D.}~\bibnamefont
  {Ayres}} \emph {et~al.},\ }\href@noop {} {\  (\bibinfo {year} {2005})},\
  \bibinfo {note} {available at \url{hep-ex/0503053v1}}\BibitemShut {NoStop}%
\bibitem [{\citenamefont {Abe}\ \emph {et~al.}(2011{\natexlab{b}})\citenamefont
  {Abe}, \citenamefont {Abe}, \citenamefont {Aihara}, \citenamefont {Fukuda},
  \citenamefont {Hayato} \emph {et~al.}}]{hyperk}%
  \BibitemOpen
  \bibfield  {author} {\bibinfo {author} {\bibfnamefont {K.}~\bibnamefont
  {Abe}}, \bibinfo {author} {\bibfnamefont {T.}~\bibnamefont {Abe}}, \bibinfo
  {author} {\bibfnamefont {H.}~\bibnamefont {Aihara}}, \bibinfo {author}
  {\bibfnamefont {Y.}~\bibnamefont {Fukuda}}, \bibinfo {author} {\bibfnamefont
  {Y.}~\bibnamefont {Hayato}},  \emph {et~al.},\ }\href@noop {} {\  (\bibinfo
  {year} {2011}{\natexlab{b}})},\ \Eprint {http://arxiv.org/abs/1109.3262}
  {arXiv:1109.3262 [hep-ex]} \BibitemShut {NoStop}%
\bibitem [{\citenamefont {Acciarri}\ \emph {et~al.}(2016)\citenamefont
  {Acciarri} \emph {et~al.}}]{dune}%
  \BibitemOpen
  \bibfield  {author} {\bibinfo {author} {\bibfnamefont {R.}~\bibnamefont
  {Acciarri}} \emph {et~al.} (\bibinfo {collaboration} {DUNE}),\ }\href@noop {}
  {\  (\bibinfo {year} {2016})},\ \Eprint {http://arxiv.org/abs/1601.05471}
  {arXiv:1601.05471 [physics]} \BibitemShut {NoStop}%
\bibitem [{\citenamefont {Martini}\ \emph {et~al.}(2013)\citenamefont
  {Martini}, \citenamefont {Ericson},\ and\ \citenamefont
  {Chanfray}}]{Martini:2012uc}%
  \BibitemOpen
  \bibfield  {author} {\bibinfo {author} {\bibfnamefont {M.}~\bibnamefont
  {Martini}}, \bibinfo {author} {\bibfnamefont {M.}~\bibnamefont {Ericson}}, \
  and\ \bibinfo {author} {\bibfnamefont {G.}~\bibnamefont {Chanfray}},\
  }\href@noop {} {\bibfield  {journal} {\bibinfo  {journal} {Phys. Rev.}\
  }\textbf {\bibinfo {volume} {D87}},\ \bibinfo {pages} {013009} (\bibinfo
  {year} {2013})},\ \Eprint {http://arxiv.org/abs/1211.1523} {arXiv:1211.1523
  [hep-ph]} \BibitemShut {NoStop}%
\bibitem [{\citenamefont {Lalakulich}\ and\ \citenamefont
  {Mosel}(2012)}]{Lalakulich:2012hs}%
  \BibitemOpen
  \bibfield  {author} {\bibinfo {author} {\bibfnamefont {O.}~\bibnamefont
  {Lalakulich}}\ and\ \bibinfo {author} {\bibfnamefont {U.}~\bibnamefont
  {Mosel}},\ }\href@noop {} {\bibfield  {journal} {\bibinfo  {journal} {Phys.
  Rev.}\ }\textbf {\bibinfo {volume} {C86}},\ \bibinfo {pages} {054606}
  (\bibinfo {year} {2012})},\ \Eprint {http://arxiv.org/abs/1208.3678}
  {arXiv:1208.3678 [nucl-th]} \BibitemShut {NoStop}%
\bibitem [{\citenamefont {Nieves}\ \emph {et~al.}(2012)\citenamefont {Nieves},
  \citenamefont {Sanchez}, \citenamefont {Simo},\ and\ \citenamefont
  {Vacas}}]{Nieves:2012yz}%
  \BibitemOpen
  \bibfield  {author} {\bibinfo {author} {\bibfnamefont {J.}~\bibnamefont
  {Nieves}}, \bibinfo {author} {\bibfnamefont {F.}~\bibnamefont {Sanchez}},
  \bibinfo {author} {\bibfnamefont {I.~R.}\ \bibnamefont {Simo}}, \ and\
  \bibinfo {author} {\bibfnamefont {M.~J.~V.}\ \bibnamefont {Vacas}},\
  }\href@noop {} {\bibfield  {journal} {\bibinfo  {journal} {Phys. Rev.}\
  }\textbf {\bibinfo {volume} {D85}},\ \bibinfo {pages} {113008} (\bibinfo
  {year} {2012})},\ \Eprint {http://arxiv.org/abs/1204.5404} {arXiv:1204.5404
  [hep-ph]} \BibitemShut {NoStop}%
\bibitem [{\citenamefont {Day}\ and\ \citenamefont
  {McFarland}(2012)}]{day_nue_numu_2012}%
  \BibitemOpen
  \bibfield  {author} {\bibinfo {author} {\bibfnamefont {M.}~\bibnamefont
  {Day}}\ and\ \bibinfo {author} {\bibfnamefont {K.~S.}\ \bibnamefont
  {McFarland}},\ }\href {\doibase 10.1103/PhysRevD.86.053003} {\bibfield
  {journal} {\bibinfo  {journal} {Phys. Rev.}\ }\textbf {\bibinfo {volume}
  {D86}},\ \bibinfo {pages} {053003} (\bibinfo {year} {2012})},\ \Eprint
  {http://arxiv.org/abs/1206.6745} {arXiv:1206.6745 [hep-ph]} \BibitemShut
  {NoStop}%
\bibitem [{\citenamefont {Hayato}(2009)}]{neut}%
  \BibitemOpen
  \bibfield  {author} {\bibinfo {author} {\bibfnamefont {Y.}~\bibnamefont
  {Hayato}},\ }\href {http://th-www.if.uj.edu.pl/acta/vol40/abs/v40p2477.htm}
  {\bibfield  {journal} {\bibinfo  {journal} {Acta Physica Polonica B}\
  }\textbf {\bibinfo {volume} {40}},\ \bibinfo {pages} {2477} (\bibinfo {year}
  {2009})}\BibitemShut {NoStop}%
\bibitem [{\citenamefont {Wilkinson}\ \emph {et~al.}(2016)\citenamefont
  {Wilkinson} \emph {et~al.}}]{niwg_paper}%
  \BibitemOpen
  \bibfield  {author} {\bibinfo {author} {\bibfnamefont {C.}~\bibnamefont
  {Wilkinson}} \emph {et~al.},\ }\href {\doibase 10.1103/PhysRevD.93.072010}
  {\bibfield  {journal} {\bibinfo  {journal} {Phys. Rev.}\ }\textbf {\bibinfo
  {volume} {D93}},\ \bibinfo {pages} {072010} (\bibinfo {year} {2016})},\
  \Eprint {http://arxiv.org/abs/1601.05592} {arXiv:1601.05592 [hep-ex]}
  \BibitemShut {NoStop}%
\bibitem [{\citenamefont {Moniz}\ \emph {et~al.}(1971)\citenamefont {Moniz},
  \citenamefont {Sick}, \citenamefont {Whitney}, \citenamefont {Ficenec},
  \citenamefont {Kephart},\ and\ \citenamefont {Trower}}]{moniz_eA_RFG_1971}%
  \BibitemOpen
  \bibfield  {author} {\bibinfo {author} {\bibfnamefont {E.~J.}\ \bibnamefont
  {Moniz}}, \bibinfo {author} {\bibfnamefont {I.}~\bibnamefont {Sick}},
  \bibinfo {author} {\bibfnamefont {R.~R.}\ \bibnamefont {Whitney}}, \bibinfo
  {author} {\bibfnamefont {J.~R.}\ \bibnamefont {Ficenec}}, \bibinfo {author}
  {\bibfnamefont {R.~D.}\ \bibnamefont {Kephart}}, \ and\ \bibinfo {author}
  {\bibfnamefont {W.~P.}\ \bibnamefont {Trower}},\ }\href {\doibase
  10.1103/PhysRevLett.26.445} {\bibfield  {journal} {\bibinfo  {journal} {Phys.
  Rev. Lett.}\ }\textbf {\bibinfo {volume} {26}},\ \bibinfo {pages} {445}
  (\bibinfo {year} {1971})}\BibitemShut {NoStop}%
\bibitem [{\citenamefont {Furmanski}(2015)}]{andy_thesis}%
  \BibitemOpen
  \bibfield  {author} {\bibinfo {author} {\bibfnamefont {A.}~\bibnamefont
  {Furmanski}},\ }\emph {\bibinfo {title} {{Charged-current Quasi-elastic-like
  neutrino interactions at the T2K experiment}}},\ \href@noop {} {Ph.D.
  thesis},\ \bibinfo  {school} {University of Warwick} (\bibinfo {year}
  {2015})\BibitemShut {NoStop}%
\bibitem [{\citenamefont {Gran}\ \emph {et~al.}(2013)\citenamefont {Gran},
  \citenamefont {Nieves}, \citenamefont {Sanchez},\ and\ \citenamefont
  {Vicente~Vacas}}]{nievesExtension}%
  \BibitemOpen
  \bibfield  {author} {\bibinfo {author} {\bibfnamefont {R.}~\bibnamefont
  {Gran}}, \bibinfo {author} {\bibfnamefont {J.}~\bibnamefont {Nieves}},
  \bibinfo {author} {\bibfnamefont {F.}~\bibnamefont {Sanchez}}, \ and\
  \bibinfo {author} {\bibfnamefont {M.}~\bibnamefont {Vicente~Vacas}},\ }\href
  {\doibase 10.1103/PhysRevD.88.113007} {\bibfield  {journal} {\bibinfo
  {journal} {Phys. Rev.}\ }\textbf {\bibinfo {volume} {D88}},\ \bibinfo {pages}
  {113007} (\bibinfo {year} {2013})},\ \Eprint {http://arxiv.org/abs/1307.8105}
  {arXiv:1307.8105 [hep-ph]} \BibitemShut {NoStop}%
\bibitem [{\citenamefont {Bodek}\ \emph {et~al.}(2015)\citenamefont {Bodek},
  \citenamefont {Christy},\ and\ \citenamefont {Coopersmith}}]{eff-sf_proc}%
  \BibitemOpen
  \bibfield  {author} {\bibinfo {author} {\bibfnamefont {A.}~\bibnamefont
  {Bodek}}, \bibinfo {author} {\bibfnamefont {M.~E.}\ \bibnamefont {Christy}},
  \ and\ \bibinfo {author} {\bibfnamefont {B.}~\bibnamefont {Coopersmith}},\
  }\bibfield  {booktitle} {\emph {\bibinfo {booktitle} {{Proceedings, Workshop
  on Neutrino Interactions, Systematic uncertainties and near detector physics:
  Session of CETUP* 2014}}},\ }\href {\doibase 10.1063/1.4931862} {\bibfield
  {journal} {\bibinfo  {journal} {AIP Conf. Proc.}\ }\textbf {\bibinfo {volume}
  {1680}},\ \bibinfo {pages} {020003} (\bibinfo {year} {2015})},\ \Eprint
  {http://arxiv.org/abs/1409.8545} {arXiv:1409.8545 [nucl-th]} \BibitemShut
  {NoStop}%
\bibitem [{\citenamefont {Wilkinson}(2015)}]{my_thesis}%
  \BibitemOpen
  \bibfield  {author} {\bibinfo {author} {\bibfnamefont {C.}~\bibnamefont
  {Wilkinson}},\ }\emph {\bibinfo {title} {{Constraining neutrino interaction
  uncertainties for oscillation experiments}}},\ \href@noop {} {Ph.D. thesis},\
  \bibinfo  {school} {University of Sheffield} (\bibinfo {year}
  {2015})\BibitemShut {NoStop}%
\bibitem [{\citenamefont {Bodek}\ \emph {et~al.}(2011)\citenamefont {Bodek},
  \citenamefont {Budd},\ and\ \citenamefont {Christy}}]{tem}%
  \BibitemOpen
  \bibfield  {author} {\bibinfo {author} {\bibfnamefont {A.}~\bibnamefont
  {Bodek}}, \bibinfo {author} {\bibfnamefont {H.~S.}\ \bibnamefont {Budd}}, \
  and\ \bibinfo {author} {\bibfnamefont {E.}~\bibnamefont {Christy}},\ }\href
  {\doibase 10.1140/epjc/s10052-011-1726-y} {\bibfield  {journal} {\bibinfo
  {journal} {Eur. Phys. J. C}\ }\textbf {\bibinfo {volume} {71}},\ \bibinfo
  {pages} {1726} (\bibinfo {year} {2011})}\BibitemShut {NoStop}%
\bibitem [{\citenamefont {Bradford}\ \emph {et~al.}(2006)\citenamefont
  {Bradford}, \citenamefont {Bodek}, \citenamefont {Budd},\ and\ \citenamefont
  {Arrington}}]{bbba05}%
  \BibitemOpen
  \bibfield  {author} {\bibinfo {author} {\bibfnamefont {R.}~\bibnamefont
  {Bradford}}, \bibinfo {author} {\bibfnamefont {A.}~\bibnamefont {Bodek}},
  \bibinfo {author} {\bibfnamefont {H.}~\bibnamefont {Budd}}, \ and\ \bibinfo
  {author} {\bibfnamefont {J.}~\bibnamefont {Arrington}},\ }\href {\doibase
  10.1016/j.nuclphysbps.2006.08.028} {\bibfield  {journal} {\bibinfo  {journal}
  {Nuclear Physics B - Proceedings Supplements}\ }\textbf {\bibinfo {volume}
  {159}},\ \bibinfo {pages} {127 } (\bibinfo {year} {2006})}\BibitemShut
  {NoStop}%
\bibitem [{\citenamefont {Kopp}(2005)}]{numi-beam}%
  \BibitemOpen
  \bibfield  {author} {\bibinfo {author} {\bibfnamefont {S.~E.}\ \bibnamefont
  {Kopp}},\ }\href@noop {} {\  (\bibinfo {year} {2005})},\ \Eprint
  {http://arxiv.org/abs/physics/0508001} {arXiv:physics/0508001 [physics]}
  \BibitemShut {NoStop}%
\bibitem [{\citenamefont {Aguilar-Arevalo}\ \emph
  {et~al.}(2009{\natexlab{b}})\citenamefont {Aguilar-Arevalo} \emph
  {et~al.}}]{mb-flux}%
  \BibitemOpen
  \bibfield  {author} {\bibinfo {author} {\bibfnamefont {A.}~\bibnamefont
  {Aguilar-Arevalo}} \emph {et~al.} (\bibinfo {collaboration} {MiniBooNE
  Collaboration}),\ }\href {\doibase 10.1103/PhysRevD.79.072002} {\bibfield
  {journal} {\bibinfo  {journal} {Phys. Rev.}\ }\textbf {\bibinfo {volume}
  {D79}},\ \bibinfo {pages} {072002} (\bibinfo {year} {2009}{\natexlab{b}})},\
  \Eprint {http://arxiv.org/abs/0806.1449} {arXiv:0806.1449 [hep-ex]}
  \BibitemShut {NoStop}%
\bibitem [{\citenamefont {Andreopoulos}\ \emph {et~al.}(2010)\citenamefont
  {Andreopoulos}, \citenamefont {Bell}, \citenamefont {Bhattacharya},
  \citenamefont {Cavanna}, \citenamefont {Dobson} \emph {et~al.}}]{genieMC}%
  \BibitemOpen
  \bibfield  {author} {\bibinfo {author} {\bibfnamefont {C.}~\bibnamefont
  {Andreopoulos}}, \bibinfo {author} {\bibfnamefont {A.}~\bibnamefont {Bell}},
  \bibinfo {author} {\bibfnamefont {D.}~\bibnamefont {Bhattacharya}}, \bibinfo
  {author} {\bibfnamefont {F.}~\bibnamefont {Cavanna}}, \bibinfo {author}
  {\bibfnamefont {J.}~\bibnamefont {Dobson}},  \emph {et~al.},\ }\href
  {\doibase 10.1016/j.nima.2009.12.009} {\bibfield  {journal} {\bibinfo
  {journal} {Nucl. Instrum. Meth.}\ }\textbf {\bibinfo {volume} {A614}},\
  \bibinfo {pages} {87} (\bibinfo {year} {2010})},\ \Eprint
  {http://arxiv.org/abs/0905.2517} {arXiv:0905.2517 [hep-ph]} \BibitemShut
  {NoStop}%
\bibitem [{Note1()}]{Note1}%
  \BibitemOpen
  \bibinfo {note} {The results used here correspond to the results with a flux
  estimate~\cite {new-minerva-flux} updated from the original publication,
  which predicts a significantly smaller flux and a smaller fractional flux
  uncertainty}\BibitemShut {NoStop}%
\bibitem [{\citenamefont {Aliaga}\ \emph {et~al.}(2016)\citenamefont {Aliaga},
  \citenamefont {Kordosky}, \citenamefont {Golan} \emph
  {et~al.}}]{new-minerva-flux}%
  \BibitemOpen
  \bibfield  {author} {\bibinfo {author} {\bibfnamefont {L.}~\bibnamefont
  {Aliaga}}, \bibinfo {author} {\bibfnamefont {M.}~\bibnamefont {Kordosky}},
  \bibinfo {author} {\bibfnamefont {T.}~\bibnamefont {Golan}},  \emph {et~al.}
  (\bibinfo {collaboration} {MINERvA Collaboration}),\ }\href@noop {} {\enquote
  {\bibinfo {title} {Private communication of manuscript in preparation,
  prediction of the numi neutrino flux},}\ } (\bibinfo {year}
  {2016})\BibitemShut {NoStop}%
\end{thebibliography}%

\clearpage

\section{Appendix: Equality of the nuclear correction for neutrinos and
  antineutrinos in the Fermi Gas model}
\label{appendix:nu-nubar-equality}

\begin{figure}[htbp]
  \centering
  \begin{subfigure}{0.45\textwidth}
    \includegraphics[width=\textwidth]{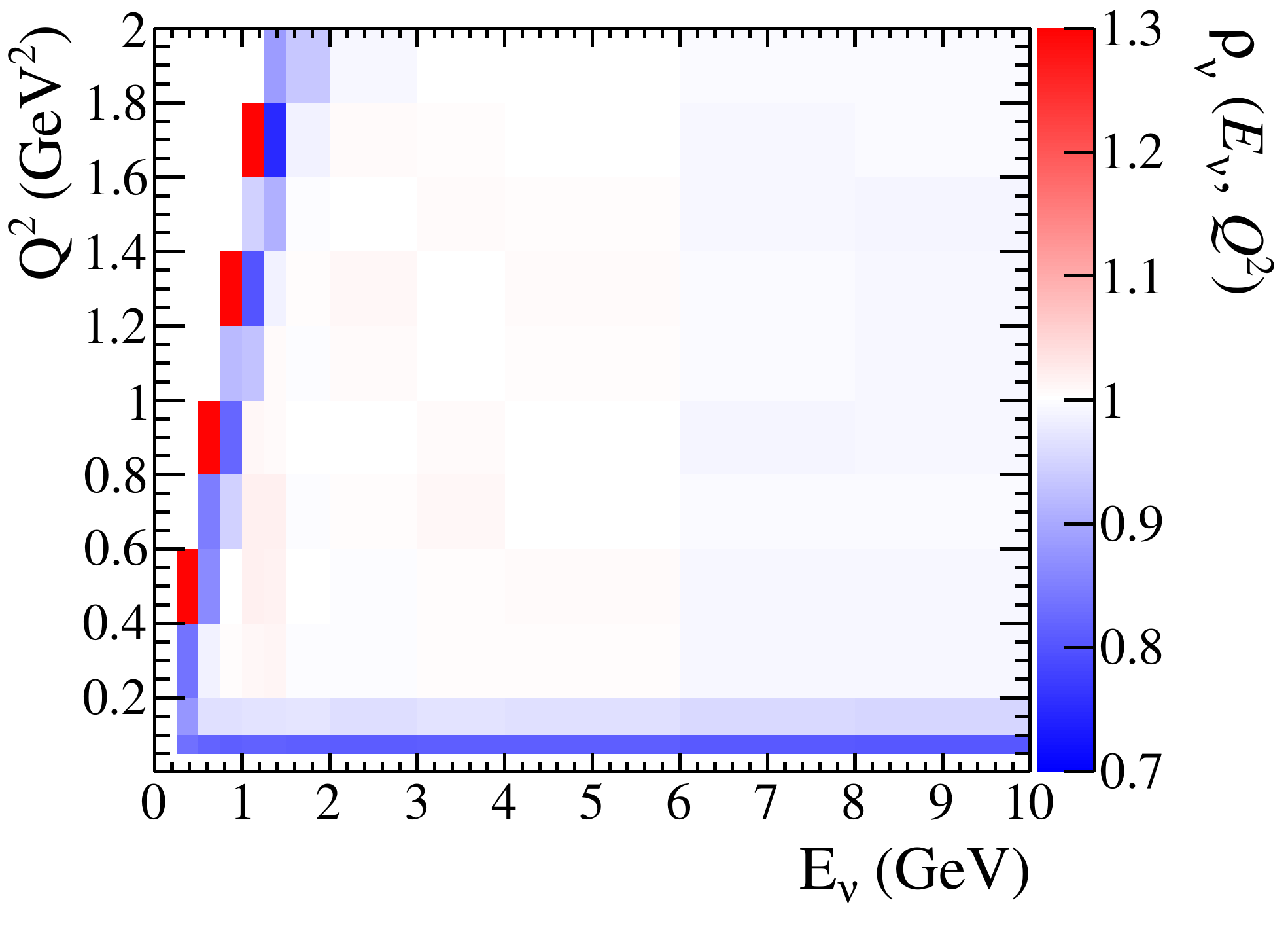}
    \caption{Neutrino}
  \end{subfigure}
  \begin{subfigure}{0.45\textwidth}
    \includegraphics[width=\textwidth]{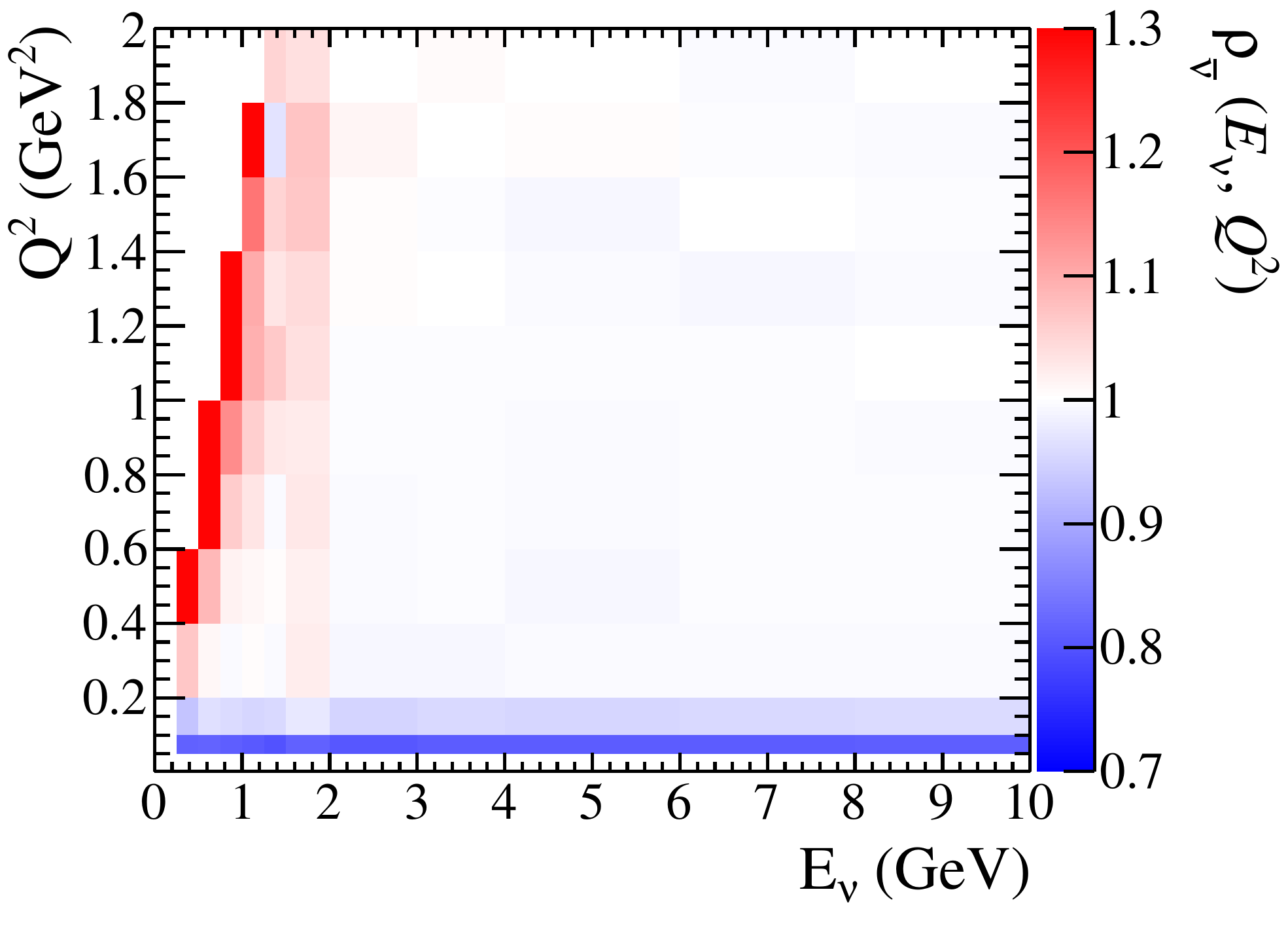}
    \caption{Antineutrino}
    \end{subfigure}
  \caption{Ratios of $\sigma(\Enu, \qq)$ for the RFG and L-S models, for both neutrino and antineutrino modes.}\label{fig:single_ratio}
\end{figure}

\begin{figure}[htbp]
  \centering
  \includegraphics[width=0.45\textwidth]{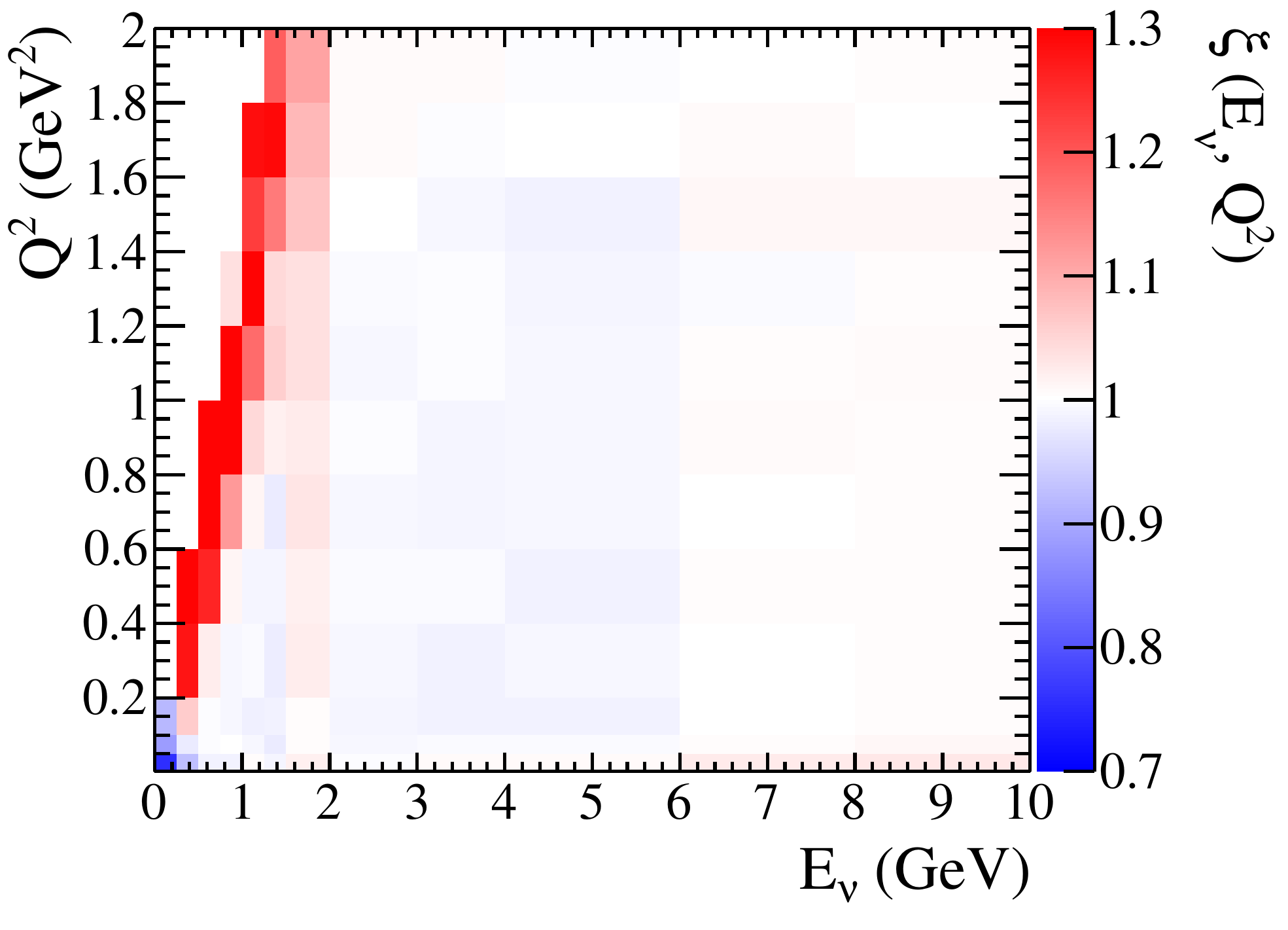}
  \caption{The double ratio $\xi(\Enu, \qq)$, defined in Equation~\ref{eq:double_ratio}, is shown. Deviations from unity indicates a bias in the technique in that region of $(\Enu, \qq)$ space.}\label{fig:double_ratio}
\end{figure}

The validity of Equation~\ref{eq:ts} rests on the
assumption that the ratio of bound to free cross sections is the
same for neutrino and antineutrino modes. 
Figure~\ref{fig:single_ratio} shows the ratio of bound to free \ccqe
cross sections for both neutrino ($\rho_{\nu}(\Enu, \qq) =
\sigma^{\mathrm{RFG}}_{\nu}(\Enu,
\qq)/\sigma^{\mathrm{free}}_{\nu}(\Enu, \qq)$) and antineutrinos
($\rho_{\bar{\nu}}(\Enu, \qq) =
\sigma^{\mathrm{RFG}}_{\bar{\nu}}(\Enu,
\qq)/\sigma^{\mathrm{free}}_{\bar{\nu}}(\Enu, \qq)$) assuming
the RFG model in GENIE for bound nucleons as a function of \Enu and
\qq. The simulated events used to produce
Figure~\ref{fig:single_ratio} are flat in neutrino energy. In
Figure~\ref{fig:double_ratio}, the double ratio
\begin{equation}
  \xi(\Enu, \qq) = \frac{\sigma^{\mathrm{RFG}}_{\bar{\nu}}(\Enu, \qq)/\sigma^{\mathrm{RFG}}_{\nu}(\Enu, \qq)}{\sigma^{\mathrm{free}}_{\bar{\nu}}(\Enu, \qq)/\sigma^{\mathrm{free}}_{\nu}(\Enu, \qq)}
  \label{eq:double_ratio}
\end{equation}
\noindent is shown, which is a direct test of this assumption for the
case of the RFG model. It can be observed from
Figure~\ref{fig:single_ratio} that at the fringe of the kinematically
allowed region, where Fermi motion increases the allowed phase space
for the RFG model, the ratio of bound to free cross sections changes
rapidly. It is clear from Figure~\ref{fig:double_ratio} that this
change is different for neutrino and antineutrino modes. This implies
that there will be a bias in the test statistic defined in
Equation~\ref{eq:ts} for neutrino energies which cannot
populate all \qq bins. \minerva, where the flux has neutrino energies
in the range $1.5 \leq \Enu \leq 10$ GeV, will not be affected by the
bias. However, \mb, with neutrino energies of $0 \leq \Enu \leq 3$
GeV, will be affected, although the size of this bias on the test
statistic is not clear from Figure~\ref{fig:double_ratio}. The biases 
are shown for both \minerva and \mb in Figure~\ref{fig:model_comparisons}.

Note that the $R(\qq)$ defined in Equation~\ref{eq:r} is the flux integrated $1/\xi(E_\nu,\qq)$ for the case of the RFG model.

\clearpage
\section{Appendix: Relationship between Measured \qqqe and \qq}
\label{appendix:Q2QE}

\begin{figure}[htbp]
  \centering
  \begin{subfigure}{0.45\textwidth}
    \includegraphics[width=\textwidth]{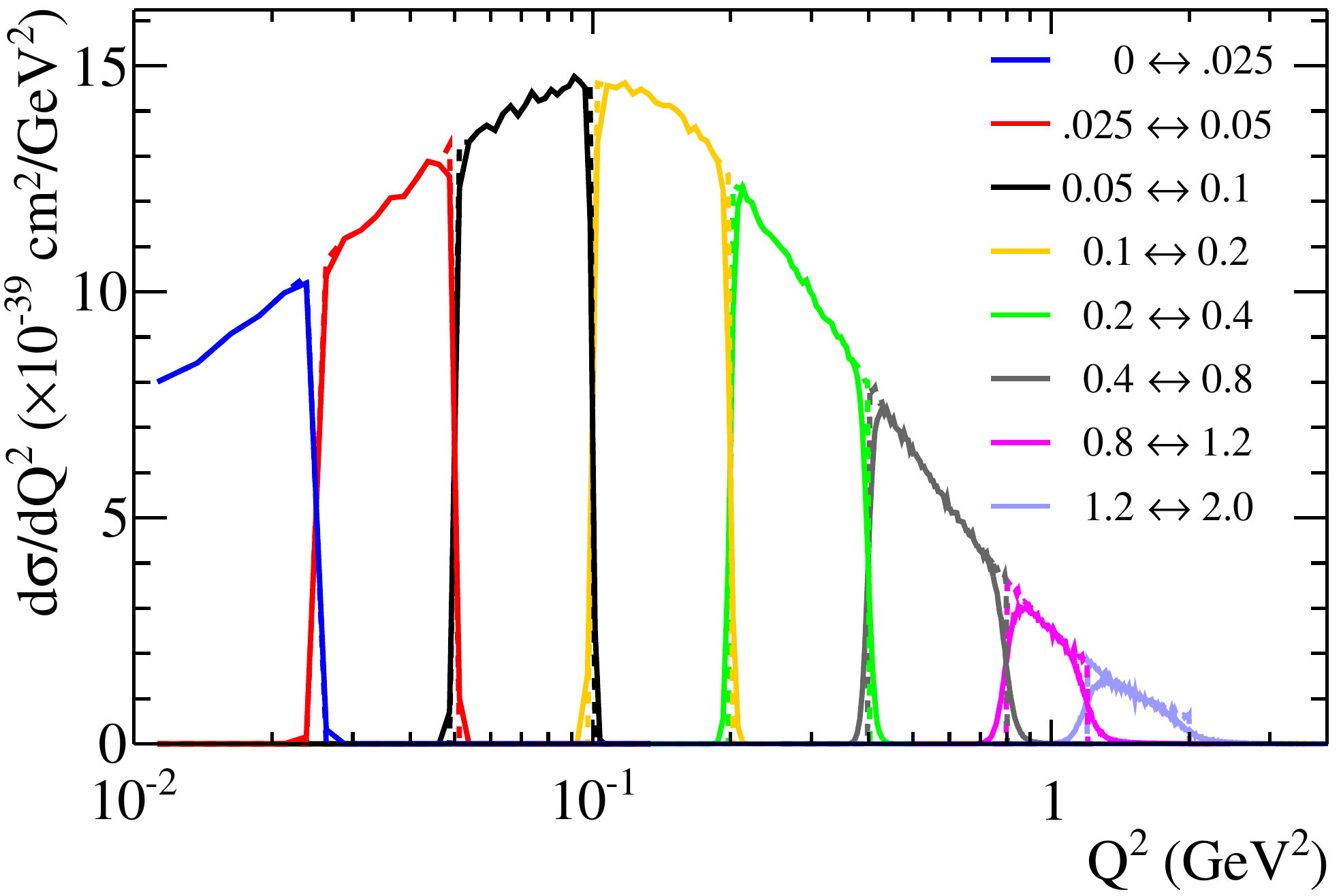}
    \caption{\minerva $\nu_{\mu}$--CH}
    \end{subfigure}
  \begin{subfigure}{0.45\textwidth}
    \includegraphics[width=\textwidth]{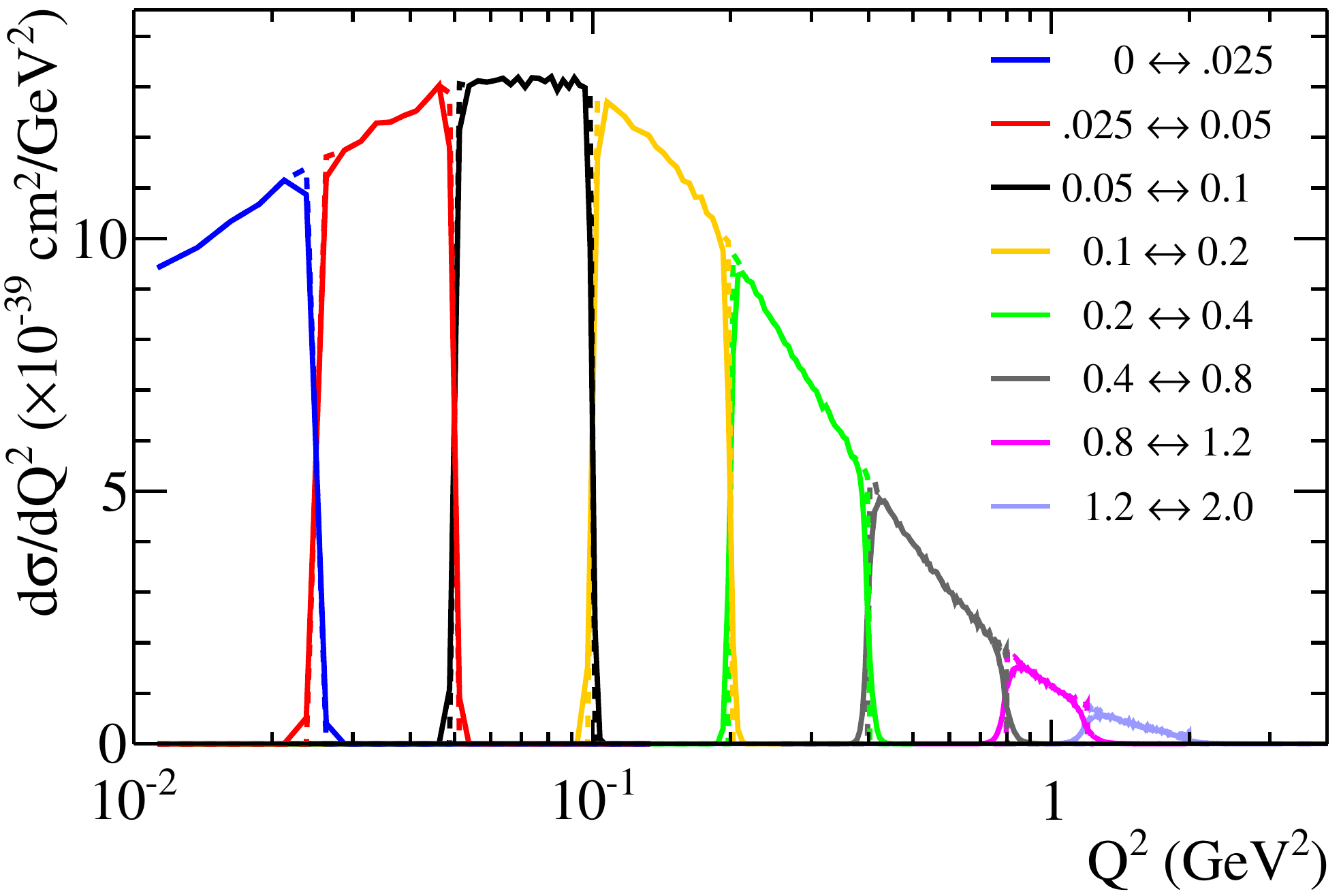}
    \caption{\minerva $\bar{\nu}_{\mu}$--CH}
    \end{subfigure}
  \caption{The $\qq \rightarrow \qqqe$ smearing is shown for the \minerva neutrino and antineutrino samples. The legend gives the \qqqe bin edges used by \minerva. The dashed lines give the flux-averaged cross section prediction for the FG model calculated using NEUT as a function of \qqqe (broken down into the \minerva binning). The solid lines show the true $\qq$ distribution of events in each \qqqe bin.}\label{fig:Q2QE_smearing_MINERvA}
\end{figure}

\begin{figure}[htbp]
  \centering
  \begin{subfigure}{0.45\textwidth}
    \includegraphics[width=\textwidth]{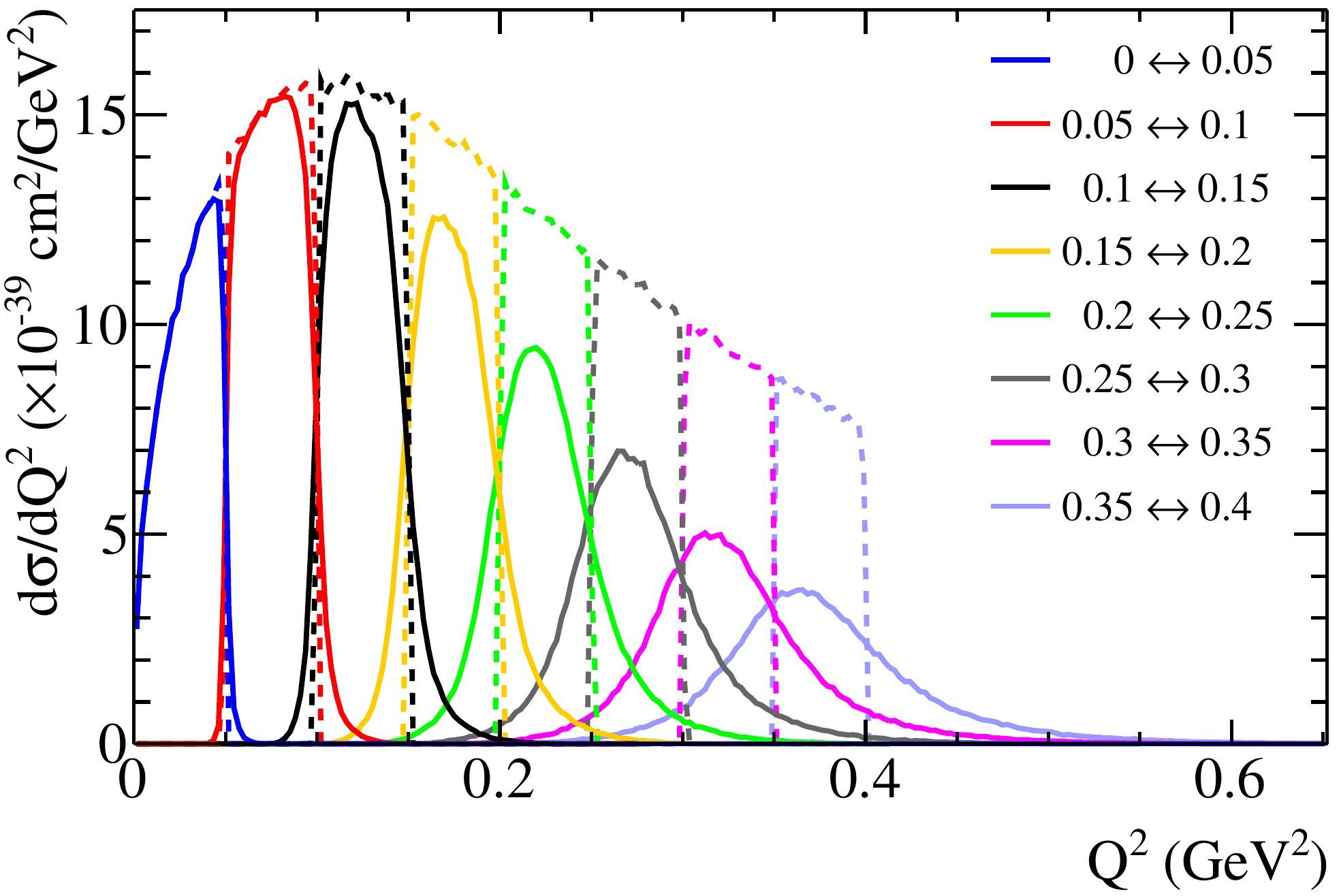}
    \caption{\mb $\nu_{\mu}$--CH$_{2}$}
    \end{subfigure}
  \begin{subfigure}{0.45\textwidth}
    \includegraphics[width=\textwidth]{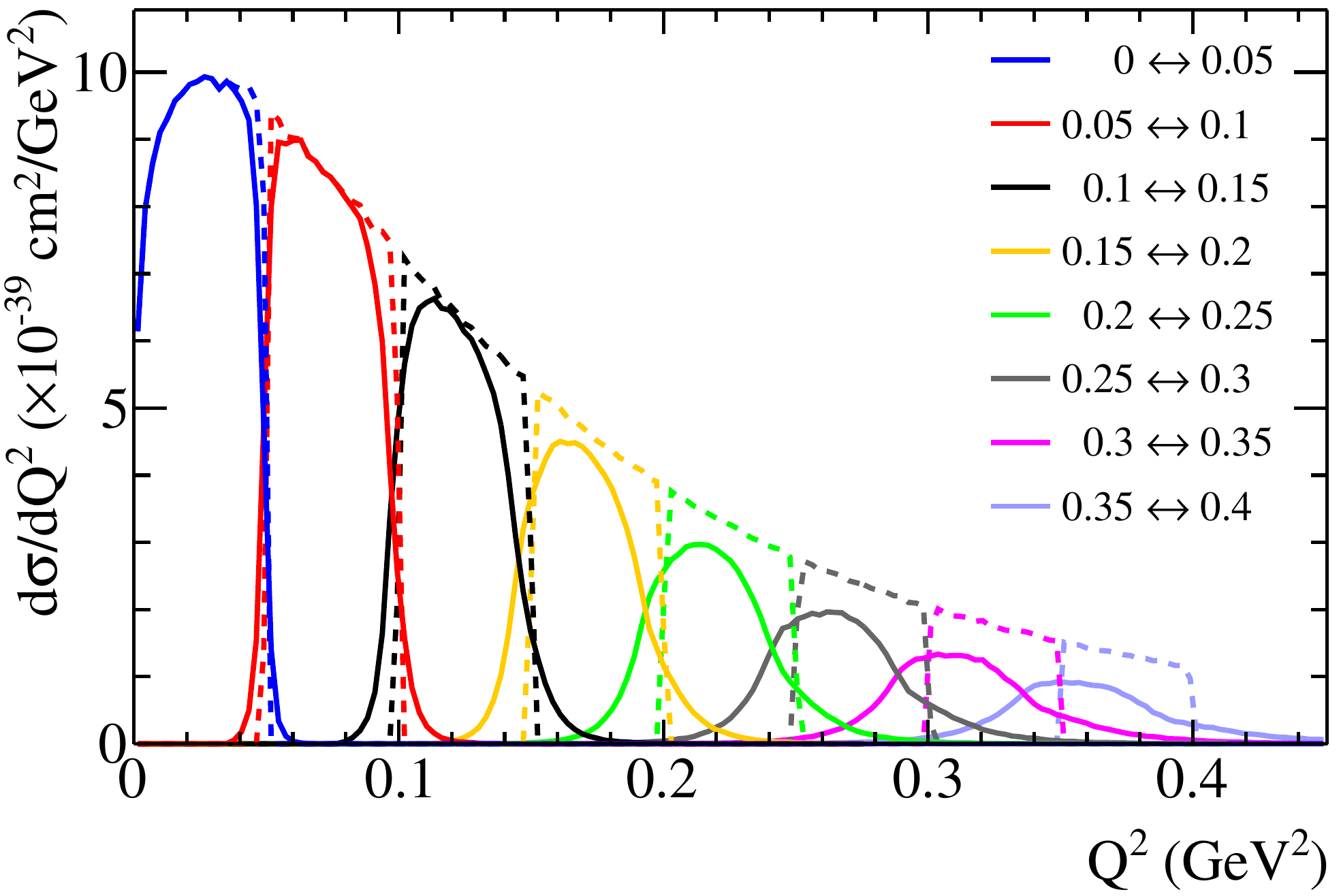}
    \caption{\mb $\bar{\nu}_{\mu}$--CH$_{2}$}
    \end{subfigure}
  \caption{The $\qq \rightarrow \qqqe$ smearing is shown for the first 8 bins of the \mb neutrino and antineutrino samples. The legend gives the \qqqe bin edges used by \mb. The dashed lines give the flux-averaged cross section prediction for the FG model calculated using NEUT as a function of \qqqe (broken down into the \mb binning). The solid lines show the true $\qq$ distribution of events in each \qqqe bin.}\label{fig:Q2QE_smearing_MiniBooNE}
\end{figure}

The \qq $\rightarrow$ \qqqe effect for the FG model is illustrated in Figure~\ref{fig:Q2QE_smearing_MINERvA} for \minerva, and Figure~\ref{fig:Q2QE_smearing_MiniBooNE} for \mb. In both figures, the true \qq distribution is shown for events which populate each of the first 8 \qqqe bins of the experiments using events simulated using the FG model in \neut with default model parameters. The smearing is not very significant for \minerva, and is minimal in the lowest \qqqe bins. For \mb, the smearing becomes significant in the higher \qqqe bins (and this trend continues for the other bins not shown in Figure~\ref{fig:Q2QE_smearing_MiniBooNE}), but is minimal at low \qqqe. \qqqe is effectively an additional smearing effect on the \qq distribution measured by the experiments, which is dependent on the nuclear model. As such it is part of the measurement of nuclear effects, but it will smear the bias introduced by correcting for the antineutrino/neutrino cross section difference with the L-S model. This effect is not corrected for, but is included in the bias tests shown on Figure~\ref{fig:model_comparisons}. Again, it is reassuring that the \qqqe smearing is minimal at low \qq.

\clearpage
\section{Appendix: Applying the method to an arbitrary theoretical model}
\label{appendix:applying}
\begin{figure*}[htbp]
  \centering
  \begin{subfigure}{0.45\textwidth}
    \includegraphics[width=\textwidth]{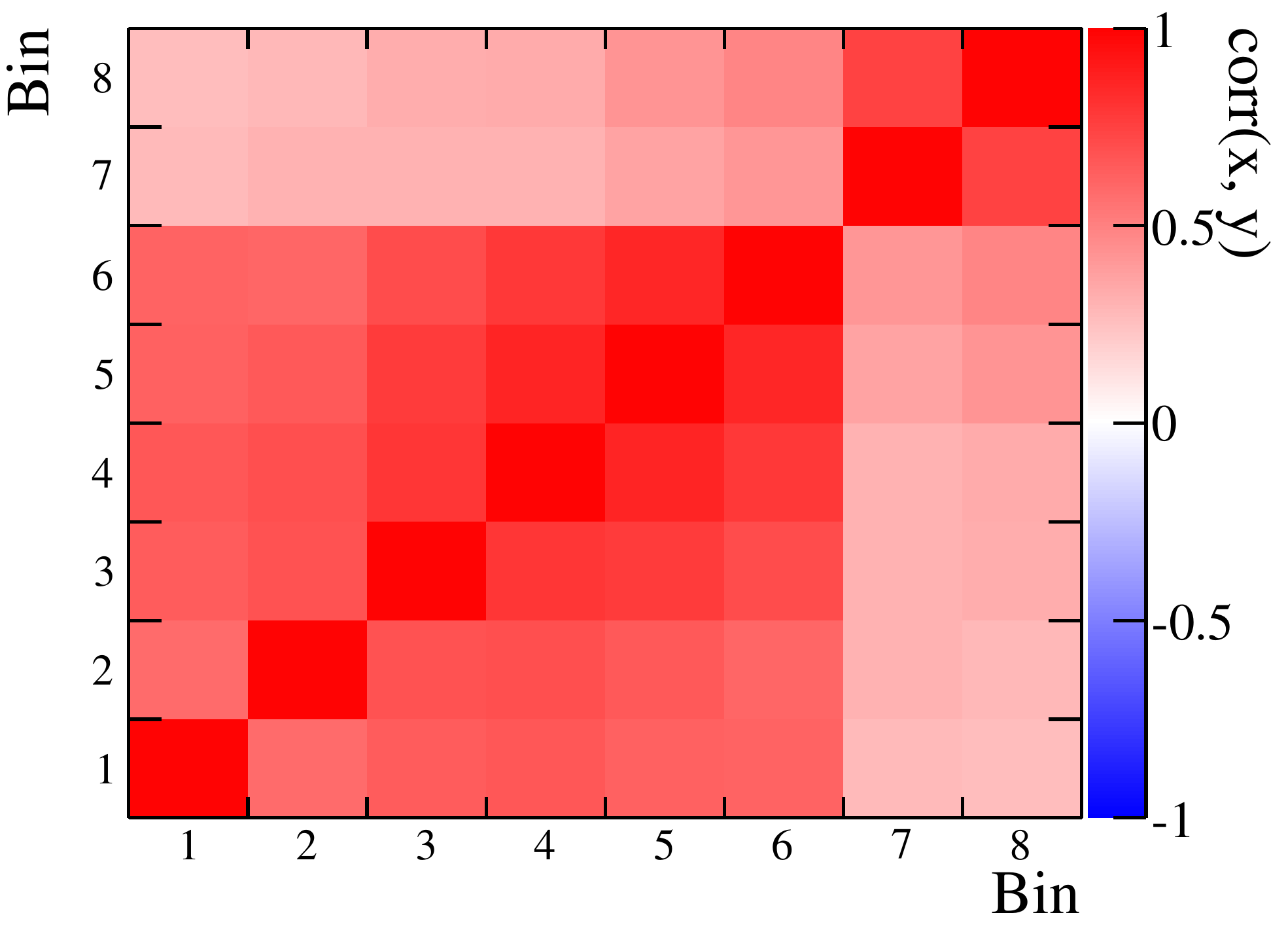}
    \caption{\minerva}
  \end{subfigure}
  \begin{subfigure}{0.45\textwidth}
    \includegraphics[width=\textwidth]{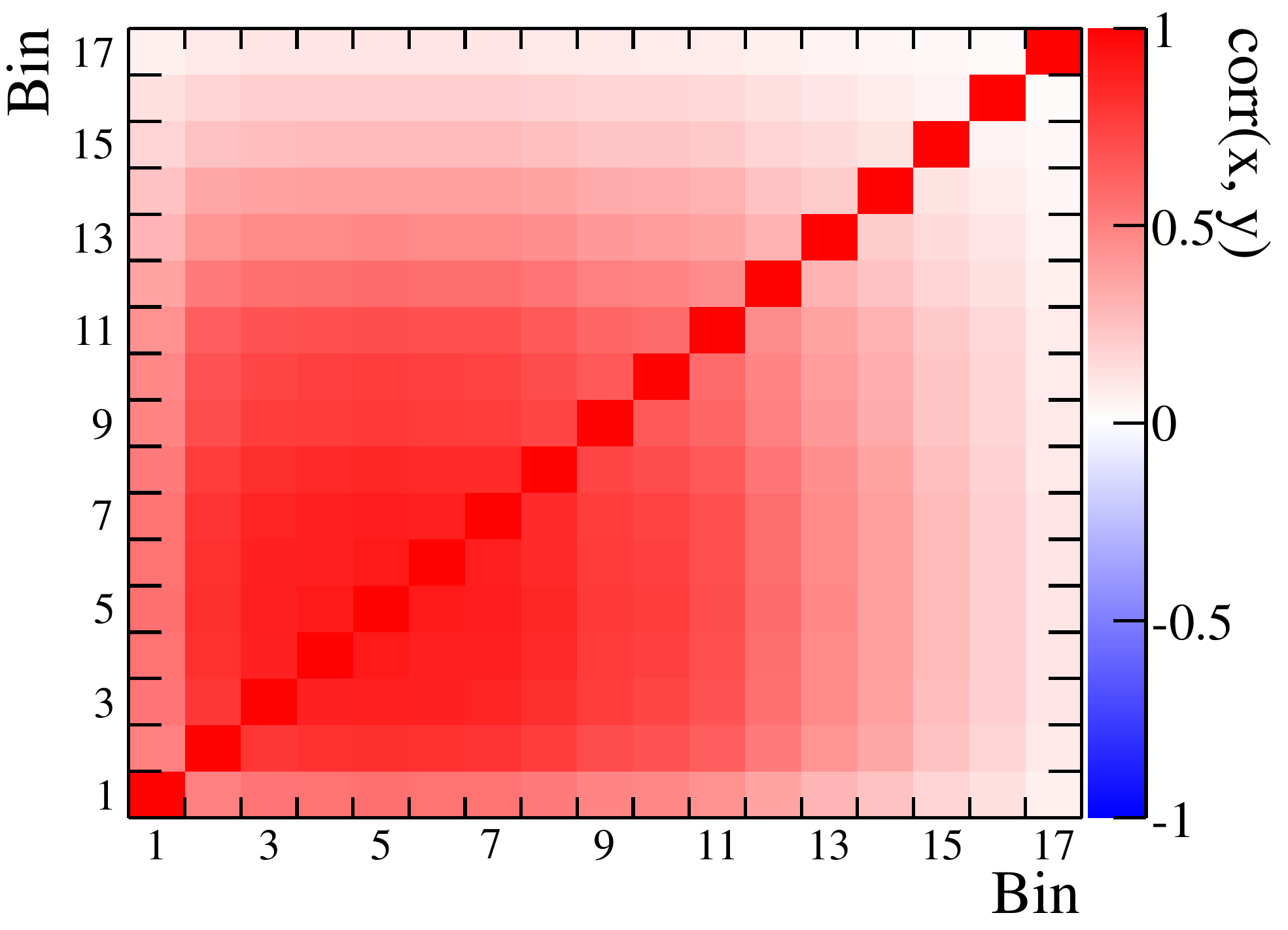}
    \caption{\mb}
  \end{subfigure}
  \caption{Correlation matrices between the measurement of the test statistic extracted from data for both \minerva and \mb. The bins numbers correspond to the increasing $\qqqe$ bins used by the experiments. The covariance matrices are given in Tables~\ref{tab:MINERvA_covar} and~\ref{tab:MiniBooNE_covar} for \minerva and \mb respectively.}\label{fig:correlation_matrices}
\end{figure*}

The extracted central values, $\lambda(\qq) = \sigma^{\bar{\nu}}_{p}(\qq)/\sigma^{\nu}_{n}(\qq)$ correction factors and covariance matrices are given for \minerva and \mb in Tables~\ref{tab:MINERvA_covar} and~\ref{tab:MiniBooNE_covar} respectively. The extracted correlation matrices are also shown for both \minerva and \mb in Figure~\ref{fig:correlation_matrices}. Note that no covariance matrix between the \mb bins has been released for either the neutrino or the antineutrino CCQE results; the correlations shown are due to the overall normalization uncertainties given independently for the neutrino (10.7\%) and antineutrino (13.0\%) data which are fully correlated between bins (but are not correlated with each other).

It is possible to apply the method outlined here to any cross section model using Equation~\ref{eq:ts}, using the $\lambda(\qq)$ correction factor. As shown in Figure~\ref{fig:model_comparisons}, the bias on the test statistic can be shown for any given model by calculating $6\sigma^{\bar{\nu}}_{\mathrm{H}}/\sigma_{\mathrm{C}}^{\bar{\nu}}$ using the test statistic defined in this work, and exactly using that model.

It is possible to form a $\chi^{2}$ statistic comparing an arbitrary model to the measurements of nuclear effects extracted here as described in Equation~\ref{eq:chi2}.
\onecolumngrid
\begin{table*}[htbp]
\small
\centering
{\renewcommand{\arraystretch}{1.2}
\begin{tabular}{c|cccccccc}
\hline\hline
\qqqe (GeV$^{2}$) bins & 0 -- 0.025  & 0.025 -- 0.05  & 0.05 -- 0.1  & 0.1 -- 0.2  & 0.2 -- 0.4  & 0.4 -- 0.8  & 0.8 -- 1.2  & 1.2 -- 2  \\ 
Test statistic & 1.61  & 0.83  & 0.85  & 0.22  & 1.06  & 0.89  & 1.66  & 2.49  \\ 
\hline
$\lambda(Q^{2})$ & 0.988  & 0.953  & 0.904  & 0.831  & 0.728  & 0.598  & 0.470  & 0.354  \\ 
\hline
0 -- 0.025 & 0.439  & 0.213  & 0.212  & 0.197  & 0.233  & 0.254  & 0.293  & 0.389  \\ 
0.025 -- 0.05 & 0.213  & 0.306  & 0.186  & 0.172  & 0.204  & 0.210  & 0.275  & 0.356  \\ 
0.05 -- 0.1 & 0.212  & 0.186  & 0.244  & 0.177  & 0.216  & 0.217  & 0.242  & 0.356  \\ 
0.1 -- 0.2 & 0.197  & 0.172  & 0.177  & 0.201  & 0.218  & 0.219  & 0.221  & 0.331  \\ 
0.2 -- 0.4 & 0.233  & 0.204  & 0.216  & 0.218  & 0.318  & 0.302  & 0.330  & 0.532  \\ 
0.4 -- 0.8 & 0.254  & 0.210  & 0.217  & 0.219  & 0.302  & 0.388  & 0.423  & 0.677  \\ 
0.8 -- 1.2 & 0.293  & 0.275  & 0.242  & 0.221  & 0.330  & 0.423  & 2.619  & 2.699  \\ 
1.2 -- 2 & 0.389  & 0.356  & 0.356  & 0.331  & 0.532  & 0.677  & 2.699  & 4.947  \\ 
\hline\hline
\end{tabular}}
\caption{The measurement of nuclear effects on carbon using \minerva
  data on $CH$, calculated using Equation~\ref{eq:ts}, and the covariance matrix between the data points.}\label{tab:MINERvA_covar}
\end{table*}

\begin{turnpage}
\begin{table*}[h!]
\centering
\footnotesize
{\renewcommand{\arraystretch}{1.2}
\begin{tabular}{c|ccccccccccccccccc}
\hline\hline
\qqqe (GeV$^{2}$) bins & 0 -- 0.05  & 0.05 -- 0.1  & 0.1 -- 0.15  & 0.15 -- 0.2  & 0.2 -- 0.25  & 0.25 -- 0.3  & 0.3 -- 0.35  & 0.35 -- 0.4  & 0.4 -- 0.45  & 0.45 -- 0.5  & 0.5 -- 0.6  & 0.6 -- 0.7  & 0.7 -- 0.8  & 0.8 -- 1  & 1 -- 1.2  & 1.2 -- 1.5  & 1.5 -- 2  \\ 
\hline
Test statistic & 2.49  & 1.59  & 1.27  & 1.29  & 1.39  & 1.33  & 1.22  & 1.39  & 1.45  & 1.61  & 1.64  & 1.42  & 1.4  & 1.38  & 0.93  & -0.31  & -1.57  \\ 
$\lambda(\qq)$ & 0.784  & 0.543  & 0.408  & 0.321  & 0.262  & 0.217  & 0.186  & 0.162  & 0.141  & 0.125  & 0.108  & 0.090  & 0.077  & 0.064  & 0.052  & 0.043  & 0.037 \\
\hline
0 -- 0.05 & 2.093  & 0.609  & 0.568  & 0.570  & 0.583  & 0.574  & 0.561  & 0.583  & 0.591  & 0.613  & 0.616  & 0.587  & 0.585  & 0.582  & 0.522  & 0.358  & 0.190  \\ 
0.05 -- 0.1 & 0.609  & 0.688  & 0.475  & 0.477  & 0.487  & 0.480  & 0.469  & 0.487  & 0.495  & 0.512  & 0.515  & 0.491  & 0.489  & 0.486  & 0.436  & 0.299  & 0.158  \\ 
0.1 -- 0.15 & 0.568  & 0.475  & 0.516  & 0.444  & 0.454  & 0.447  & 0.437  & 0.454  & 0.461  & 0.477  & 0.480  & 0.457  & 0.455  & 0.453  & 0.407  & 0.279  & 0.148  \\ 
0.15 -- 0.2 & 0.570  & 0.477  & 0.444  & 0.501  & 0.456  & 0.449  & 0.439  & 0.456  & 0.463  & 0.479  & 0.482  & 0.459  & 0.457  & 0.455  & 0.408  & 0.280  & 0.148  \\ 
0.2 -- 0.25 & 0.583  & 0.487  & 0.454  & 0.456  & 0.511  & 0.459  & 0.449  & 0.466  & 0.473  & 0.490  & 0.493  & 0.469  & 0.468  & 0.465  & 0.417  & 0.286  & 0.152  \\ 
0.25 -- 0.3 & 0.574  & 0.480  & 0.447  & 0.449  & 0.459  & 0.509  & 0.442  & 0.459  & 0.466  & 0.483  & 0.486  & 0.463  & 0.461  & 0.458  & 0.411  & 0.282  & 0.149  \\ 
0.3 -- 0.35 & 0.561  & 0.469  & 0.437  & 0.439  & 0.449  & 0.442  & 0.495  & 0.449  & 0.455  & 0.471  & 0.474  & 0.452  & 0.450  & 0.448  & 0.402  & 0.275  & 0.146  \\ 
0.35 -- 0.4 & 0.583  & 0.487  & 0.454  & 0.456  & 0.466  & 0.459  & 0.449  & 0.588  & 0.473  & 0.490  & 0.493  & 0.469  & 0.467  & 0.465  & 0.417  & 0.286  & 0.152  \\ 
0.4 -- 0.45 & 0.591  & 0.495  & 0.461  & 0.463  & 0.473  & 0.466  & 0.455  & 0.473  & 0.712  & 0.497  & 0.500  & 0.476  & 0.474  & 0.472  & 0.424  & 0.290  & 0.154  \\ 
0.45 -- 0.5 & 0.613  & 0.512  & 0.477  & 0.479  & 0.490  & 0.483  & 0.471  & 0.490  & 0.497  & 0.813  & 0.518  & 0.493  & 0.491  & 0.489  & 0.439  & 0.301  & 0.159  \\ 
0.5 -- 0.6 & 0.616  & 0.515  & 0.480  & 0.482  & 0.493  & 0.486  & 0.474  & 0.493  & 0.500  & 0.518  & 0.952  & 0.496  & 0.494  & 0.492  & 0.441  & 0.302  & 0.160  \\ 
0.6 -- 0.7 & 0.587  & 0.491  & 0.457  & 0.459  & 0.469  & 0.463  & 0.452  & 0.469  & 0.476  & 0.493  & 0.496  & 1.270  & 0.471  & 0.468  & 0.420  & 0.288  & 0.153  \\ 
0.7 -- 0.8 & 0.585  & 0.489  & 0.455  & 0.457  & 0.468  & 0.461  & 0.450  & 0.467  & 0.474  & 0.491  & 0.494  & 0.471  & 1.892  & 0.466  & 0.419  & 0.287  & 0.152  \\ 
0.8 -- 1 & 0.582  & 0.486  & 0.453  & 0.455  & 0.465  & 0.458  & 0.448  & 0.465  & 0.472  & 0.489  & 0.492  & 0.468  & 0.466  & 2.800  & 0.416  & 0.285  & 0.151  \\ 
1 -- 1.2 & 0.522  & 0.436  & 0.407  & 0.408  & 0.417  & 0.411  & 0.402  & 0.417  & 0.424  & 0.439  & 0.441  & 0.420  & 0.419  & 0.416  & 4.462  & 0.256  & 0.136  \\ 
1.2 -- 1.5 & 0.358  & 0.299  & 0.279  & 0.280  & 0.286  & 0.282  & 0.275  & 0.286  & 0.290  & 0.301  & 0.302  & 0.288  & 0.287  & 0.285  & 0.256  & 4.152  & 0.093  \\ 
1.5 -- 2 & 0.190  & 0.158  & 0.148  & 0.148  & 0.152  & 0.149  & 0.146  & 0.152  & 0.154  & 0.159  & 0.160  & 0.153  & 0.152  & 0.151  & 0.136  & 0.093  & 4.036  \\
\hline\hline
\end{tabular}}
\caption{The measurement of nuclear effects on carbon using \mb data
  on $CH_2$, calculated using Equation~\ref{eq:ts}, and the covariance matrix between the data points.}\label{tab:MiniBooNE_covar}
\end{table*}
\end{turnpage}
\twocolumngrid

\end{document}